\documentclass{aa}  

\usepackage{graphicx}
\usepackage{txfonts}
\usepackage{hyperref}
\def\Nsubjects{288,109}

\def\mstar{M_*}

\def\zphot{z_{\mathrm{phot}}}

\def\Sref#1{Sect.~\ref{#1}\xspace}
\def\Fref#1{Fig.~\ref{#1}\xspace}
\def\Tref#1{Table~\ref{#1}\xspace}
\def\Eref#1{Eq.~\ref{#1}\xspace}

\def\pr{{\rm P}}

\def\icut{i_{\mathrm{cut}}}
\def\rcut{r_{\mathrm{cut}}}
\def\gcut{g_{\mathrm{cut}}}

\def\tein{\theta_{\mathrm{Ein}}}
\def\tsource{\theta_{s}}

\def\nimg{N_{\mathrm{img}}}
\def\nvisible{n_{\mathrm{visible}}}

\def\data{\mathbf{d}}

\def\datakt{\mathbf{d}_k^t}
\def\datakpot{\mathbf{d}_{k+1}^t}
\def\dataonet{\mathbf{d}_1^t}

\def\NgradeA{14}
\def\NgradeB{129}
\def\NgradeAB{143}
\def\NgradeC{581}
\def\NgradeCabove{724}
\def\Nknown{70}
\def\Nyatta{6779}
\def\Nprob{1577}
\def\Ntalk{264}
\def\NsugohiAB{324}
\def\nseen{N_{\mathrm{seen}}}

\def\Nmax{30}

\def\pr{{\rm P}}

\def\yatta{{\sc YattaLens}}

\setcitestyle{notesep={}}

\begin{document}

   \title{Survey of gravitationally lensed objects in HSC Imaging (SuGOHI). VI. Crowdsourced lens finding with Space Warps}
   \titlerunning{SuGOHI VI.}
   \authorrunning{Sonnenfeld et al.}

   \author{Alessandro Sonnenfeld\inst{\ref{leiden},\ref{ipmu}}\thanks{Marie Sk\l{}odowska-Curie Fellow}\and
          Aprajita Verma\inst{\ref{oxford}}\and
          Anupreeta More\inst{\ref{ipmu},\ref{iucaa}}\and
          Elisabeth Baeten\inst{\ref{zoo}}\and
          Christine Macmillan\inst{\ref{zoo}}\and
          Kenneth C. Wong\inst{\ref{ipmu},\ref{naoj}}\and
          James~H.~H.~Chan\inst{\ref{epfl}}\and
          Anton T. Jaelani\inst{\ref{kindai},\ref{bandung}}\and
          Chien-Hsiu Lee\inst{\ref{noao}}\and
          Masamune Oguri\inst{\ref{ipmu},\ref{todai},\ref{masamune3}}\and
          Cristian E. Rusu\inst{\ref{subaru}}\and
          Marten Veldthuis\inst{\ref{oxford}}\and
          Laura Trouille\inst{\ref{adler}}\and
          Philip J. Marshall\inst{\ref{kipac}}\and
          Roger Hutchings\inst{\ref{oxford}}\and
          Campbell Allen\inst{\ref{oxford}}\and
          James O'~Donnell\inst{\ref{oxford}}\and
          %z2
          Claude Cornen\inst{\ref{zoo}}\and
          Christopher P. Davis\inst{\ref{kipac}}\and
          Adam McMaster\inst{\ref{oxford}}\and
          %z3
          Chris Lintott\inst{\ref{oxford}}\and
          Grant Miller\inst{\ref{oxford}}
          }

   \institute{Leiden Observatory, Leiden University, Niels Bohrweg 2, 2333 CA Leiden, the Netherlands\\
              \email{sonnenfeld@strw.leidenuniv.nl}\label{leiden}
             \and
            Kavli IPMU (WPI), UTIAS, The University of Tokyo, Kashiwa, Chiba 277-8583, Japan\label{ipmu} \and
            Sub-department of Astrophysics, University of Oxford, Denys Wilkinson Building, Keble Road, Oxford OX1 3RH, UK\label{oxford}\and
            The Inter-University Center for Astronomy and Astrophysics, Post bag 4, Ganeshkhind, Pune, 411007, India\label{iucaa}\and
            Zooniverse, c/o Astrophysics Department, University of Oxford, Oxford\label{zoo}\and
            Institute of Physics, Laboratory of Astrophysics, \'{E}cole Polytechnique F\'{e}d\'{e}rale de Lausanne (EPFL), Observatoire de Sauverny, 1290 Versoix, Switzerland\label{epfl}\and
            Department of Physics, Kindai University, 3-4-1 Kowakae, Higashi-Osaka, Osaka 577-8502, Japan\label{kindai}\and
            Astronomy Study Program and Bosscha Observatory, FMIPA, Institut Teknologi Bandung, Jl. Ganesha 10, Bandung 40132, Indonesia\label{bandung}\and
            National Optical Astronomy Observatory 950 N Cherry Avenue, Tucson, AZ 85719, USA\label{noao}\and
            Kavli Institute for Particle Astrophysics and Cosmology, Stanford University, 452 Lomita Mall, Stanford, CA 94035, USA\label{kipac}\and
            Department of Physics, University of Tokyo, 7-3-1 Hongo, Bunkyo-ku, Tokyo 113-0033, Japan\label{todai}\and
            Research Center for the Early Universe, University of Tokyo, 7-3-1 Hongo, Bunkyo-ku, Tokyo 113-0033, Japan\label{masamune3}\and
            Subaru Telescope, National Astronomical Observatory of Japan, 2-21-1 Osawa, Mitaka, Tokyo 181-0015, Japan\label{subaru}\and
            National Astronomical Observatory of Japan, 2-21-1 Osawa, Mitaka, Tokyo 181-8588, Japan\label{naoj}\and
            Adler Planetarium, Chicago, IL, 60605, USA\label{adler}
             }

   \date{}

% \abstract{}{}{}{}{} 
% 5 {} token are mandatory
 
  \abstract
   {Strong lenses are extremely useful probes of the distribution of matter on galaxy and cluster scales at cosmological distances, however, they are rare and difficult to find. The number of currently known lenses is on the order of 1000.}
   {The aim of this study is to use crowdsourcing to carry out a lens search targeting massive galaxies selected from over 442 square degrees of photometric data from the Hyper Suprime-Cam (HSC) survey.}
   {
Based on the S16A internal data release of the HSC survey, we chose a sample of $\sim300000$ galaxies with photometric redshifts in the range of $0.2 < \zphot < 1.2$ and photometrically inferred stellar masses of $\log{\mstar} > 11.2$. 
We crowdsourced lens finding on this sample of galaxies on the Zooniverse platform as part of the Space Warps project. The sample was complemented by a large set of simulated lenses and visually selected non-lenses for training purposes.
Nearly $6000$ citizen volunteers participated in the experiment.
In parallel, we used \yatta,\, an automated lens-finding algorithm, to look for lenses in the same sample of galaxies.
}
   {
Based on a statistical analysis of classification data from the volunteers, we selected a sample of the most promising $\sim1,500$ candidates, which we then visually inspected: half of them turned out to be possible (grade C) lenses or better.
By including lenses found by \yatta\ or serendipitously noticed in the discussion section of the Space Warps website, we were able to find $\NgradeA$ definite lenses (grade A), $\NgradeB$ probable lenses (grade B), and $\NgradeC$ possible lenses. \yatta\ found half the number of lenses that were  discovered via crowdsourcing.}
   {
Crowdsourcing is able to produce samples of lens candidates with high completeness, when multiple images are clearly detected, and with higher purity compared to the currently available automated algorithms.
A hybrid approach, in which the visual inspection of samples of lens candidates pre-selected by discovery algorithms or coupled to machine learning is crowdsourced, will be a viable option for lens finding in the 2020s, with forthcoming wide-area surveys such as LSST, Euclid, and WFIRST.
}

   \keywords{Gravitational lensing: strong --
             Galaxies: elliptical and lenticular, cD
               }

   \maketitle
%
%________________________________________________________________
\section{Introduction}\label{sect:intro}

Strong gravitational lensing is a very powerful tool for studying galaxy evolution and cosmology.
For example, strong lenses have been used to explore the inner structure of galaxies and their evolution \citep[e.g.][]{T+K02, K+T03, Aug++10, Ruf++11, Bol++12, Son++13b}, 
as well as to put constraints on the stellar initial mass function (IMF) of massive galaxies \citep[e.g.][]{Tre++10, Spi++12, Bar++13, SLC15, Son++19b} and on their dark matter content \citep[e.g.][]{Son++12, New++15, O+A18}. Strong lensing is a unique tool for detecting the presence of substructures inside or along the line of sight of massive galaxies \citep[e.g.][]{M+S98, Mor++09, Veg++10, Hsu++20}. Strongly lensed compact sources, such as quasars or supernovae, have been used to measure the surface mass density in stellar objects via the microlensing effect \citep[e.g.][]{Med++09, Sch++14, ORF14} and to measure cosmological parameters from time delay observations \citep[e.g.][]{Suy++17, Gri++18, Won++20, Mil++20}.
While they are very useful, strong lenses are rare, as they require the chance alignment of a light source with a foreground object with a sufficiently large surface mass density.
The number of currently known strong lenses is on the order of 1000, with the exact number depending on the purity of the sample\footnote{More than half of the systems considered for this estimate are candidates with high probability of being lenses, but no spectroscopic confirmation.}.
Despite this seemingly high number, the effective sample size is, in practice, much smaller for many strong lensing applications once the selection criteria for obtaining suitable objects for a given study are applied. 
%For example, strongly lensed quasars are only a small fraction of the full sample of lenses. 
For example, most known lens galaxies have redshifts of $z<0.5$, limiting the time range that can be explored in evolution studies.
For this reason, many strong lensing-based inferences are still dominated by statistical uncertainties due to small sample sizes.
Expanding the sample of known lenses would, therefore, broaden the range of investigations that can be carried out, providing statistical power that is presently lacking.

Current wide-field photometric surveys, such as the Hyper Suprime-Cam Subaru Strategic Program \citep[HSC SSP][]{Aih++18,Miy++18}, the Kilo Degree Survey \citep[KiDS][]{deJ++15}, and the Dark Energy Survey \citep[DES][]{DES16} are allowing the discovery of hundreds of new promising strong lens candidates \citep[e.g.][]{Son++18, Pet++19, Jac++19}.
Although the details vary between surveys, the general strategy for finding new lenses consists of scanning images of galaxies that exhibit the potential of serving as lenses, given their mass and redshift, and looking for the presence of strongly lensed features around them.
Due to the large areas covered by the aforementioned surveys (the HSC SSP is planned to acquire data over 1400 square degrees of sky), the number of galaxies that are to be analysed in order to obtain a lens sample as complete as possible can easily reach into the hundreds of thousands. 
In order to deal with such a large scale, the lens finding task is usually automated, either by making use of a lens finding algorithm or artificial neural networks trained on simulated data \citep[see][ for an overview of some of the latest methods employed for lens finding in purely photometric data]{Met++19}.
We point out how the current implementations of automatic lens finding algorithms, including those based on artificial neural networks, require some degree of visual inspection: typically these methods are applied in such a way as to prioritise completeness, resulting in a relatively low purity. For example, out of 1480 lens candidates found by the algorithm \yatta\ in the HSC data, only 46 were labelled as highly probable lens candidates \citep{Son++18}. Similarly, the convolutional neural networks developed by \citet{Pet++19} for a lens search in KiDS data produced a list of 3,500 candidates, of which only 89 were recognised to be strong lenses with high confidence after visual inspection.
Nevertheless, \citet{Pet++19} showed how high purity can still be achieved without human intervention, albeit with a great loss in completeness.

While it is both desirable and plausible that future improvements in the development of lens finding algorithms will lead to higher purity and completeness in lens searches,
a currently viable and very powerful approach to lens finding is crowdsourcing, which harnesses the skill and adaptability of human pattern recognition. 
With crowdsourcing, lens candidates are distributed among a large number of trained volunteers for visual inspection. The Space Warps collaboration has been pioneering this method and has applied it successfully to data from the Canada-France-Hawaii Telescope Legacy Survey \citep[CFHT-LS][]{Mar++16, Mor++16}.
In this work, we use crowdsourcing 
and the tools developed by the Space Warps team
to look for strong lenses in 442 square degrees of imaging data collected from the HSC survey.

We obtained cutouts around $\sim300000$ massive galaxies selected with the criteria listed above and submitted them for inspection to a team of citizen scientist volunteers, together with training images consisting of known lenses, simulated lenses, and non-lens galaxies, via the Space Warps platform.
The volunteers were asked to simply label each image as either a lens or a non-lens.
After collecting multiple classifications for each galaxy in the sample, we combined them in a Bayesian framework to obtain a probability for an object to be a strong lens. The science team then visually inspected the most likely lens candidates and classified them with more stringent criteria.
In parallel to crowdsourcing, we searched for strong lenses in the same sample of massive galaxies by using the  \yatta\ software, which has been used for past lens searches in HSC data \citep{Son++18, Won++18}.
By merging the crowdsourced lens candidates with those obtained by \yatta, we were able to discover a sample of $\NgradeAB$ high-probability lens candidates. Most of these very promising candidates were successfully identified as such by citizen participants.

The aim of this paper is to describe the details of our lens finding effort, present the sample of newly discovered lens candidates, discuss the relative performance of crowdsourcing and of the \yatta\ software, and to suggest strategies for future searches.
The structure of this work is as follows.
In \Sref{sect:data}, we describe the data used for the lens search, including the training set for crowdsourcing.
In \Sref{sect:crowd}, we describe the setup of the crowdsourcing experiment and the algorithm used for the analysis of the classifications from the citizen scientist volunteers.
In \Sref{sect:results}, we show our results, including the sample of candidates found with \yatta\ and highlighting interesting lens candidates of different types.
In \Sref{sect:discuss}, we discuss the merits and limitations of the two lens finding strategies. We conclude in \Sref{sect:concl}.
All magnitudes are in AB units and all images are oriented with north up and east to the left.

%__________________________________________________________________

\section{The data}\label{sect:data}

\subsection{The parent sample}\label{ssec:sample}

Our lens-finding effort was based on a targeted search, as opposed to a blind one: we looked for lensed features among a set of galaxies based on the properties that make them potential lenses. 
Specifically, we targeted galaxies in the redshift range of $0.2 < z < 1.2$ and with stellar mass $M_*$ larger than $10^{11.2}M_\odot$, with photometric data from the HSC survey.
The upper redshift and lower stellar mass bounds are a compromise between the goal of obtaining a sample of lenses as complete as possible and the need to keep the number of galaxies to be inspected by the volunteers within a reasonable value,
while the lower redshift cut was introduced to avoid dealing with galaxies too bright for the detection of lensed features, which are typically faint.
In order to select such a sample, we relied on the photometric redshift catalogue from the S16A internal data release of the HSC survey \citep{Tan++18}. In particular, we used data products obtained with the template fitting-based photometric redshift software {\sc Mizuki} \citep{Tan++15}, which fits, explicitly and self-consistently, the star-formation history of a galaxy and its redshift, using the five bands of HSC ($g, r, i, z, y$).
We applied the redshift and stellar mass cuts listed above to the median value of the photometric redshift and the stellar mass provided by {\sc Mizuki}, for each galaxy with photometric data in all five HSC bands and detections in at least three bands, regardless of depth.
We removed galaxies with saturated pixels, as well as probable stars, by setting {\tt i\_extendedness\_value}~$>0$ and by removing objects brighter than 21~mag in the $i-$band and a moments-derived size smaller than $0.4''$. Typical statistical uncertainties are $0.02$ on the photo-z and $0.05$ on $\log{M_*}$.

The steps described above led us to a sample of $\sim300000$ galaxies.
From these, we removed $\Nknown$ known lenses from the literature, mostly from our previous searches \citep{Son++18, Won++18, Cha++20}, as well as a few hundred galaxies that have already been inspected and identified as possible lenses (grade C candidates in our notation) in the aforementioned studies. Many of the known lenses were used for training purposes, which is further explained in Subsection \ref{ssec:training}.
The \citet{Son++18} search covered the same area as the present study, but targeted exclusively $\sim43000$ luminous red galaxies with spectra from the Baryonic Oscillation Spectroscopic Survey.
Only about half of those galaxies belong to the sample used for our study, while the remaining half was excluded because of our stellar mass limit.
Finally, we removed from the sample $\sim4000$ galaxies used as negative (i.e. non-lens) training subjects\footnote{The term subject refers to cutouts centred on our target galaxies.}. The selection of these objects is described in Subsection \ref{ssec:training}.
The final sample size consisted of $\Nsubjects$ subjects.

Although our goal is to maximise completeness, some lenses were inevitably missed in this initial step. These are lenses for which the deflector is outside the redshift range considered or below the stellar mass cut. Additionally, blending between lens and source light can be responsible for inaccuracies and, in the worst cases, catastrophic failures in the determination of the lens photometric redshift or stellar mass, which compromises completeness. Lenses with a relatively bright source and small image separation are particularly affected by this and, thus, likely to be missing from our sample.

\subsection{HSC imaging}

We used $g,r,i$-band data from HSC S17A\footnote{S17A is a more recent data release than S16A, on which the target selection was based. While reduced data from S17A were available at the start of our experiment, the photo-z catalogue was not, hence we used the S16A photo-z catalogue to define the sample of targets.} for our experiment.
The target depth of the Wide layer of the HSC SSP, where our targets are located, is $26.2$, $26.4$ and $26.8$ in $i-$, $r-$ and $g-$band respectively \citep[$5\sigma$ detection for a point source][]{Aih++18b}. Roughly 70\% of our objects lie in parts of the survey for which the target depth has been reached in S17A.
The median seeing in the $i-$, $r-$ and $g-$band is $0.57''$, $0.67''$ and $0.72''$ , respectively \citep{Aih++18b}.
Image quality and depth are critical for the purpose of finding lenses. 
Compared to KiDS and DES, HSC data is both sharper and deeper and should therefore allow the discovery of lenses that are fainter or have smaller image separation, with respect to the typical lenses found in these other surveys (for more, see our discussion in Subsection \ref{ssec:depth}).

For each subject, we obtained $101\times101$ pixel (i.e. $17.0\times17.0''$) cutouts from co-added and sky-subtracted data in each band, which we used both for our crowdsourced search and for the search with \yatta. These data, among with corresponding variance maps and models of the point spread function (PSF), were produced by the HSC data reduction pipeline {\sc HSCPipe} version 5.4 \citep{Bos++18}, a version of the Large Synoptic Survey Telescope stack \citep{Ive++08, Axe++10, Jur++15}.

\subsection{RGB Image preparation}

In order to facilitate the visual detection of faint lensed features that would normally be blended with the lens light, we produced versions of the images of the subjects with the light from the main (foreground, putative lens) galaxy subtracted off. Foreground light subtraction was carried out by fitting a S\'{e}rsic surface brightness profile to the data, using \yatta. The structural parameters of the S\'{e}rsic model (e.g. the half-light radius, position angle, etc.) were first optimised on the $i-$band data (the band with the best image quality), then a scaled-up version of the model, convolved with the model PSF, was subtracted from the data in each band.
The presence of lens light-subtracted images was one of the main elements of novelty in our experiment, compared to past searches with Space Warps, which, in fact, recommended the adoption of such a procedure to improve the detection of lenses.

\begin{figure}
\includegraphics[width=\columnwidth]{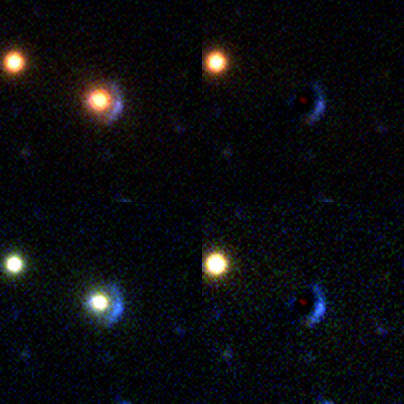}
\caption{Colour-composite HSC images of lens SL2SJ021411$-$040502. Images with the light from the foreground galaxy subtracted are shown on the right, while original images are on the left. The images at the top and bottom row were created with the `standard' and `optimal' colour schemes, respectively.
\label{fig:known}
}
\end{figure}

The original and foreground-subtracted data were used to make two sets of RGB images, with different colour schemes.
All colour schemes were based on a linear mapping between the flux in each pixel and the intensity in the 8-bit RGB channel of the corresponding band:
\begin{equation}
\begin{split}
& R = 255\times \rm{min}\left(i/\icut, 1\right),\\
& G = 255\times \rm{min}\left(r/\rcut, 1\right),\\
& B = 255\times \rm{min}\left(g/\gcut, 1\right).\\
\end{split}
\end{equation}
In the above equations, $i,r,g$ indicate the flux in each band, while $i_{\mathrm{cut}}$, $r_{\mathrm{cut}}$ and $g_{\mathrm{cut}}$ are threshold values: pixels with higher flux than these thresholds are assigned the maximum allowed intensity, 255. In a similar vein, pixels with negative flux are given 0 intensity.
In the first colour scheme, dubbed `standard', we fixed $\icut$, $\rcut$ and $\gcut$ for all images, set to sufficiently low values so as to make it possible to detect objects with surface brightness close to the background noise level.
This choice resulted in images with consistent colours for different targets, though often with saturated centres, especially for the brightest galaxies.
We also adopted a second colour scheme, named `optimal', where we used a dynamical definition of the threshold flux for each subject: in the original image (i.e. without foreground subtraction), this was set to the 99\%-ile of the pixel values in each band, so that the main galaxy, typically the brightest object in the cutout, was not saturated, except for the very central pixels.
The `optimal' threshold flux of the foreground-subtracted images was instead set to the minimum between the 90\%-ile of the residual image and the flux corresponding to ten times the sky background fluctuation. We made this choice to highlight features close to the background noise level.
As an example, we show the aforementioned four versions of the RGB images of a known lens, SL2SJ021411$-$040502, in \Fref{fig:known}.

\section{Crowdsourcing experiment setup}\label{sect:crowd}

Our crowdsourcing project, named Space Warps - HSC, is hosted on the Zooniverse\footnote{\url{https://www.zooniverse.org}} platform.
The setup of the experiment followed largely that of previous Space Warps efforts, with minor modifications. We summarise it here and refer to \citet{Mar++16} for further details.

Upon landing on the Space Warps website\footnote{\url{spacewarps.org}}, volunteers were presented with two main options: reading a brief documentation on the basic concepts of gravitational lensing or moving directly to the image classification phase. The documentation includes various examples of typical strong lens configurations, as well as false positives: non-lens galaxies with lens-like features such as spiral arms or star-forming rings and typical image artefacts.
During the image classification phase, volunteers were shown sets of four images of individual subjects, of the kind shown in \Fref{fig:known}, and asked to decide whether the subject showed any signs of gravitational lensing, in which case they were asked to click on the image, or to proceed to the next subject.
On the side of the classification interface, a `Field Guide' with a summary of various lens types and common impostors was always available for volunteers to consult.
Users accessing the image classification interface for the first time were guided through a brief tutorial, which summarised the goals of the crowdsourcing experiment, the basics of gravitational lensing, and the classification submission procedure.

In addition to the documentation, the `Field Guide' and the tutorial, we relied on training images to help volunteers sharpen their classification skills. 
Participants were shown subjects, to be graded, interleaved with training images of lenses (known or simulated), known as `sims' for simplicity, or of non-lens galaxies, referred to as `duds'.
They were not told whether a subject was a training one until after they submitted their classification, when a positive or negative feedback message was displayed, depending on whether they guessed the correct nature of the subject (lens or non-lens) or not (with some exceptions, described later).
Training images were interleaved in the classification stream with a frequency of one in three for the first 15 subjects shown, reducing to one in five for the next 25 subjects and then settling to a rate of one in ten as volunteers became more experienced. The sims and duds were randomly served throughout the experiment to each registered volunteer. As the number of sims was 50\% higher than the duds in the training subject pool, the sims were shown with correspondingly higher frequency than the duds.
We describe in detail the properties of the training images in Subsection \ref{ssec:training}.

Volunteers also had the opportunity to discuss the nature of individual subjects on the `Talk' (i.e. forum) section of the Space Warps website: after deciding the class of a subject, clicking on the `Done \& Talk' button would submit the classification and prompt the volunteer to a dedicated forum page, where they could leave comments, ask questions, and view any previous related discussion.
Volunteers did not have the possibility of changing their classification once it was submitted, so the main purposes of this forum tool was to give them a chance to bring the attention on specific subjects and ask for opinions from other volunteers or experts. This helped the volunteers in building a better understanding of gravitational lensing, as well as creating a sense of community.
We regularly browsed `Talk' to answer questions and to look for outstanding subjects highlighted by volunteers.

Volunteer classifications were compiled and analysed using the Space Warps Analysis Pipeline \citep[{\sc swap}]{Mar++16} to obtain probabilities of each subject being a lens. We describe {\sc swap} in Subsection \ref{ssec:swap}, and in practice use a modified version of the implementation of {\sc swap} written for the Zooniverse platform by Michael Laraia et al.\footnote{The modified {\sc SWAP-2} branch used here can be found at \url{https://github.com/cpadavis/swap-2} which is based on \url{https://github.com/miclaraia/swap-2}}.

\subsection{The training sample}\label{ssec:training}

Training subjects served three different purposes. The first was helping volunteers to learn how to distinguish lenses from non-lenses, as discussed above. The second purpose was to keep volunteers alert through pop-up boxes that give real-time feedback on their classifications of the training images: given that the fraction of galaxies that are strong lenses in real data is very low, showing long uninterrupted sequences of subjects could have led volunteers to adopt a default non-lens classification, which could have resulted in the mis-classification of the rare, but extremely valuable, real lenses.
The third purpose was allowing us to evaluate the accuracy of the classifications by volunteers, so as to adjust the weight of their scores in the calculation of the lens probabilities of subjects (more details to follow in Subsection \ref{ssec:swap}).
In order to serve these functions properly, it was crucial for training subjects to be as similar as possible to real ones.
This required having a large number of them, so that volunteers could always be shown training images that had never been seen previously\footnote{In practice, due to some platform/image server issues, some volunteers saw a small fraction of training subjects more than once. However, only the first classification made by a user of any given subject was used in SWAP.}.
We prepared images of thousands of training subjects across two classes: lenses and non-lens impostors.

\subsubsection{The lens sample}\label{sssec:sims}

Lens-training subjects were, for the most part, simulated ones, generated by adding images of strongly lensed galaxies on top of HSC images of galaxies from the Baryon Acoustic Oscillations Survey \citep[BOSS,][]{Daw++13} luminous red galaxy samples.
Our priority was to generate simulations covering as large a volume of parameter space as possible, within the realm of galaxy-scale lenses, so as not to create a bias against rare lens configurations.
For this reason, rather than assuming a physical model, we imposed very loose conditions on the mapping between the observed properties of the galaxies selected to act as lenses and their mass distribution.
Given a BOSS galaxy, we first assigned a lens mass profile to it in the form of a singular isothermal ellipsoid \citep[SIE,][]{KSB94}.
We drew the value of the Einstein radius $\tein$ from a Gaussian distribution centred on $\tein=1.5''$, with a dispersion  of $0.5''$, and truncated below $0.5''$ and above $3.0''$. The lower limit was set to roughly match the resolution limit of HSC data (the typical $i-$band seeing is $0.6''$), while the upper limit was imposed to restrict the simulations to galaxy-scale lenses (as opposed to group- or cluster-scale lenses, which have typical Einstein radii of several arcseconds).
We drew the lens mass centroid from a uniform distribution within a circle of one pixel radius, centred on the central pixel of the cutout (which typically coincides with the galaxy light centroid).
We drew the axis ratio of the SIE from a uniform distribution between $0.4$ and $1.0$, while the orientation of the major axis was drawn from a Normal distribution centred on the lens galaxy light major axis and with a 10~degree dispersion.

The background source was modelled with an elliptical S\'{e}rsic light distribution. 
Its half-light radius, S\'{e}rsic index and axis ratio were drawn from uniform distributions in the ranges $0.2''-3.0''$, $0.5-2.0$ and $0.4-1.0$, respectively, and its position angle was randomly oriented.
We assigned intrinsic (i.e. unlensed) source magnitudes in $g,r,i$ bands to those of objects randomly drawn from the CFHTLenS photometric catalogue \citep{Hil++12,Erb++13}.
Although the CFHTLenS survey has comparable depth to HSC, magnification by lensing makes it possible to detect sources that are fainter than the nominal survey limits, but these faint sources are missing from our training sample. Nevertheless, for extended sources such as strongly lensed images, surface brightness is more important than magnitude in determining detectability.
Since we assigned source half-light radii independently of magnitudes, our source sample spanned a broad range in surface brightness, including sources that are below the sky background fluctuation level.

The source position distribution was drawn from the following axi-symmetric distribution:
\begin{equation}\label{eq:rdist}
\pr(\tsource) \propto \left(\frac{\tsource}{\tein}\right)\exp{\left\{-4\frac{\tsource}{\tein}\right\}},
\end{equation}
where $\tsource$ is the radial distance between the source centroid and the centre of the image.
The above distribution is approximately linear in $\tsource$ at small radii, as one would expect for sources uniformly distributed in the plane of the sky, but then peaks at $\tein/4$ and turns off exponentially at large radii.
The rationale for this choice was to down-weight the number of lenses with a very asymmetric image configuration, which correspond to values of $\tsource$ that are close to the radial caustic of the SIE lens (i.e. the largest allowed value of $\tsource$ for a source to be strongly lensed), at an angular scale $\approx\tein$.
Sources close to the radial caustic are lensed into a main image, subject to minimal lensing distortion, and a very faint (usually practically invisible) counter-image close to the centre. These systems are strong lenses from a formal point of view, but in practice, they are hard to identify as such. They would dominate the simulated lens population if we assumed a strictly uniform spatial distribution of sources, hence the alternative distribution of \Eref{eq:rdist}.

Not all lens-source pairs generated this way were strong lenses: in $\sim13\%$ of cases the source fell outside the radial caustic. Such systems were simply removed from the sample.
Among the remaining simulations, most showed clear signatures of strong lensing (e.g. multiple images or arcs). For some, however, it was difficult to identify them as lenses from visual inspection alone.
We decided to include in the training sample all strong lenses, regardless of how obvious their lens nature was, so that volunteers would have the opportunity to learn how to identify lenses in the broadest possible range of configurations. This choice could also allow us to carry out a quantitative analysis of the completeness of crowdsourced lens finding as a function of a variety of lens parameters, although that is beyond the scope of this work.

We split the lens training sample in two categories: an `easy' subsample, consisting of objects showing fairly obvious strong lens features, and a `hard' subsample, consisting of less trivial lenses to identify visually.
After classifying an easy lens, volunteers received a positive feedback message (``Congratulations! You spotted a simulated lens'') or a negative one (``Oops, you missed a simulated lens. Better luck next time!''), depending on whether they correctly identified it as a lens or not.
For hard lenses, we used a different feedback rule: a correct identification still triggered a positive feedback message (``Congratulations! You've found a simulated lens. This was a tough one to spot, well done for finding it.''), but no feedback message was provided in case of misidentification, in order not to discourage volunteers with unrealistic expectations (often the lensed images in these hard sims were impossible to see at all). The implementation of two levels of feedback is a novelty of this study, compared to previous Space Warps experiments.

\begin{figure}
\includegraphics[width=\columnwidth]{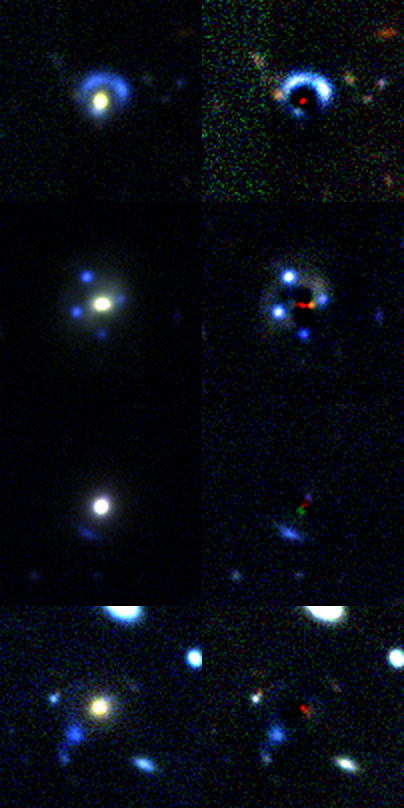}
\caption{
Set of four simulated lenses, rendered using the `optimised' colour scheme, with and without foreground subtraction.
The first two lenses from the top are labelled as easy, while the bottom two are examples of hard lenses.
\label{fig:sims}
}
\end{figure}

The separation of the lens training sample in easy and hard categories was based on the following algorithm, developed in a few iterations involving the visual inspection of a small sample of simulated lenses. For each lens, we first defined the lensed source footprint as the ensemble of cutout pixels in which the source $g-$band flux exceeded the sky background noise by more than $3\sigma$.
We then counted the number of connected regions in pixel space, using the function {\sc label} from the {\sc measure} module of the Python {\sc scikit-image} package, and used it as a proxy for the number of images $\nimg$.
We also counted the number $\nvisible$ of `visible' source pixels (not necessarily connected) where the source surface brightness exceeded that of the lens galaxy in the $g-$band. The latter was estimated from the best-fit S\'{e}rsic model of the lens light.
Any subject with $\nvisible < 40$~pixels or $\nimg=0$ was labelled as a hard one. Hard lenses were also systems with $\nimg=1$ but with a source footprint smaller than 100 pixels. Among lenses with $\nimg > 1$, those with the footprints of the brightest and second brightest images smaller than 100 and 20 pixels respectively were also given a hard lens label. All other lenses were classified as easy.
We converged to these values after inspecting a sample of simulated lenses, by making sure that the classification obtained with this algorithm matched our judgement of what constitutes an easy and a hard lens.
We show examples of lenses from the two categories in \Fref{fig:sims}.
We generated a total of $\sim12000$ simulated lenses, to which we added 52 known lenses from the literature. About 60\% of them were easy lenses.

Although the BOSS galaxies used as lenses are labelled as red, a substantial fraction of them are late-type galaxies, meaning that they exhibit spiral arms, disks, or rings.
Simulations with a late-type galaxy as a lens are more difficult to recognise because the colours of the lensed images are often similar to those of star-forming regions in the lens galaxy. Nevertheless, we allowed late-type galaxies as lenses in the training sample as we did not want to bias the volunteers against this class of objects.
Similarly, we did not apply any cut aimed at eliminating simulated lenses with an unusually large Einstein radius for their stellar mass. While unlikely, the existence of lenses with a large mass-to-light ratio cannot be excluded a priori due to the presence of dark matter and we wished the volunteers to be able to recognise them if they happened to be present in the data.
In general, it is not crucial for the lens sample to closely follow  the distribution of real lenses; in terms of lens or source properties: in order to meet our training goals, the most important requirement.
 is to span the largest volume in parameter space.  %

\subsubsection{The non-lens sample}

The most difficult aspect of lens finding through visual inspection is distinguishing true lenses, which are intrinsically rare, from non-lens galaxies with lens-like features such as spiral arms or, more generally, tangentially elongated components with different colours than those of the main galaxy body.
The latter are much more common than the former, so any inaccuracy in the classification has typically a large impact on the purity of a sample of lens candidates.
In order to maximise opportunities for volunteers to learn how to differentiate between the two categories, we designed our duds training set by including exclusively non-lens objects bearing some degree of resemblance to strong lenses.

\begin{figure}
\includegraphics[width=\columnwidth]{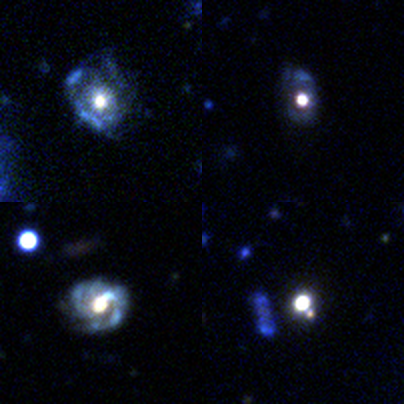}
\caption{
Set of four non-lens duds from the training set, rendered using the `optimised' colour scheme.
\label{fig:duds}
}
\end{figure}

We searched for suitable galaxies among a sample of $\sim6600$ lens candidates identified by \yatta, which we ran over the whole sample of subjects before starting the crowdsourcing experiment.
Details on the lens search with the \yatta\ algorithm are given in Subsection \ref{ssec:yatta}.
Upon visual inspection, a subset of $\sim3800$ galaxies were identified as unambiguous non-lenses and deemed suitable for training purposes. These were used as our sample of duds.
We show examples of duds in \Fref{fig:duds}.
In order to double the number of available duds, we also included in the training set versions of the original duds rotated by 180 degrees.

\subsection{The classification analysis algorithm}\label{ssec:swap}

The {\sc swap} algorithm was introduced and discussed extensively by \citet{Mar++16}. Here, we summarise its main assumptions.
The goal of the crowdsourcing experiment is to quantify for each subject the posterior probability of it being a lens based on the data, $\pr(\rm{LENS}|\data)$. The data used for the analysis consists of the ensemble of classifications from all users who have seen the subject. This includes, for the $k-$th user, the classification on the subject itself, $C_k$, as well as past classifications on training subjects $\datakt$:
\begin{equation}
\data = \left(\{C_k\},\{\datakt\}\right),
\end{equation}
where curly brackets denote ensembles over all volunteers who have classified the subject and the classification $C_k$ can take the values 'LENS' or 'NOT'.

Using Bayes' theorem, the posterior probability of a subject being a lens given the data is
\begin{equation}
\pr(\rm{LENS}|\{C_k\},\{\datakt\}) = \dfrac{\pr(\rm{LENS})\pr(\{C_k\}|\rm{LENS},\{\datakt\})}{\pr(\{C_k\}|\{\datakt\})},
\end{equation}
where $\pr(\rm{LENS})$ is the prior probability of a subject being a lens, $\pr(\{C_k\}|\rm{LENS},\{\datakt\})$ is the likelihood of obtaining the ensemble of classifications given that the subject is a lens and given the past classifications of volunteers on training subjects, while $\pr(\{C_k\}|\{\datakt\})$ is the probability of obtaining the classifications, marginalised over all possible subject classes:
\begin{equation}
\begin{split}
\pr(\{C_k\}|\{\datakt\}) = & \pr(\{C_k\}|\rm{LENS},\{\datakt\})\pr(\rm{LENS}) + \\
& \pr(\{C_k\}|\rm{NOT},\{\datakt\})\pr(\rm{NOT}).
\end{split}
\end{equation}
Before any classification takes place, the posterior probability of a subject being a lens is equal to its prior, which we assume to be
\begin{equation}
\pr(\rm{LENS}) = 2\times10^{-4},
\end{equation}
loosely based on estimates from past lens searches in CFHT-LS data.
Although our HSC data is slightly better than CFHT-LS both in terms of image quality and depth, which should correspond in principle to a higher fraction of lenses, we do not expect that to make a significant difference. This is because, as we clarify later in this paper, we designed our experiment so that the final posterior probability of a subject is always dominated by the likelihood and not by the prior.

After the first classification is made, $C_1$, we update the posterior probability, which becomes
\begin{equation}\label{eq:postone}
\pr(\rm{LENS}|C_1,\dataonet) = \dfrac{\pr(\rm{LENS})\pr(C_1|\rm{LENS},\dataonet)}{\pr(C_1§\dataonet)}.
\end{equation}
We evaluate the likelihood based on past performance of the volunteer on training subjects. We approximate the probability of the volunteer correctly classifying a lens subject with the rate at which they did so on training subjects:
\begin{equation}\label{eq:lenslike}
\pr(\rm{'LENS'}|\rm{LENS},\dataonet) \approx \frac{N_{\rm{'LENS'}}}{N_{\rm{LENS}}},
\end{equation}
where $N_{\rm{LENS}}$ is the number of sims the volunteer classified and $N_{\rm{'LENS'}}$ the number of times they classified these sims as lenses.
Given that 'LENS' and 'NOT' are the only two possible choices, the probability of the same volunteer wrongly classifying a lens as a non-lens is
\begin{equation}
\pr(\rm{'NOT'}|\rm{LENS},\dataonet) = 1 - \pr(\rm{'LENS'}|\rm{LENS},\dataonet).
\end{equation}
Similarly, we approximate the probability of a volunteer correctly classifying a dud as 
\begin{equation}\label{eq:notlike}
\pr(\rm{'NOT'}|\rm{NOT},\dataonet) \approx \frac{N_{\rm{'NOT'}}}{N_{\rm{NOT}}}.
\end{equation}

Let us now consider a subject for which $k$ classifications from an equal number of volunteers have been gathered. 
If a $k+1$-th classification is collected, we can use the posterior probability of the subject being a lens after the first $k$ classifications, $\pr(\rm{LENS}|C_1,\ldots,C_k,\dataonet,\ldots,\datakt)$, as a prior for the probability of the subject being a lens before the new classification is read. The posterior probability of the subject being a lens after the $k+1$th classification then becomes:
\begin{equation}\label{eq:recurse}
\begin{split}
& \pr(\rm{LENS}|C_{k+1},\datakpot,C_1,\ldots,C_k,\dataonet,\ldots,\datakt) = \\
& \dfrac{\pr(\rm{LENS}|C_1,\ldots,C_k,\dataonet,\ldots,\datakt)\pr(C_{k+1}|\rm{LENS},\datakpot)}{\pr(C_{k+1}|\datakpot)},
\end{split}
\end{equation}
where the probability of observing a classification $C_{k+1}$, the denominator of the above equation, is
\begin{equation}
\begin{split}
& \pr(C_{k+1}|\datakpot) = \\
& \pr(C_{k+1}|\rm{LENS},\datakpot)\pr(\rm{LENS}|C_1,\ldots,C_k,\dataonet,\ldots,\datakt) + \\
& \pr(C_{k+1}|\rm{NOT},\datakpot)\pr(\rm{NOT}|C_1,\ldots,C_k,\dataonet,\ldots,\datakt).
\end{split}
\end{equation}
\Eref{eq:recurse} allows us to update the probability of a subject being a lens every time a new classification is submitted.

As shown in past Space Warps experiments, after a small number of classifications is collected (typically 11 for a lens and 4 for a non-lens), $\pr(\rm{LENS}|\data)$ almost always converges to either very low values, indicating that the subject is most likely not a lens, or to values very close to unity, suggesting that the subject is a lens \citep[see e.g. Figure 5 of][]{Mar++16}.
The posterior probability is in either case very different from the prior, indicating that the likelihood terms are driving the inference.
In order to make the experiment more efficient, we retired subjects (i.e. we stopped showing them to the volunteers) when they reached a lens probability smaller than $10^{-5}$ after at least four classifications: gathering additional classifications would not have changed the probability of those subjects significantly, and removing them from the sample allowed us to prioritise subjects with fewer classifications.
Regardless of $\pr(\rm{LENS}|\data)$, we retired subjects after $\Nmax$ classifications were collected.
In practice, {\sc swap} was not run continuously, but only every 24 hours. This caused minor inconsistencies between the retirement rules described above and the subjects being shown to the volunteers. These inconsistencies, along with other unreported issues, such as the retirement server being offline and a delayed release of subjects, did slightly reduce the overall efficiency of the experiment, but these did not affect the probability analysis.

\section{Results}\label{sect:results}

\subsection{Search with Space Warps}

\begin{figure}
\includegraphics[width=\columnwidth]{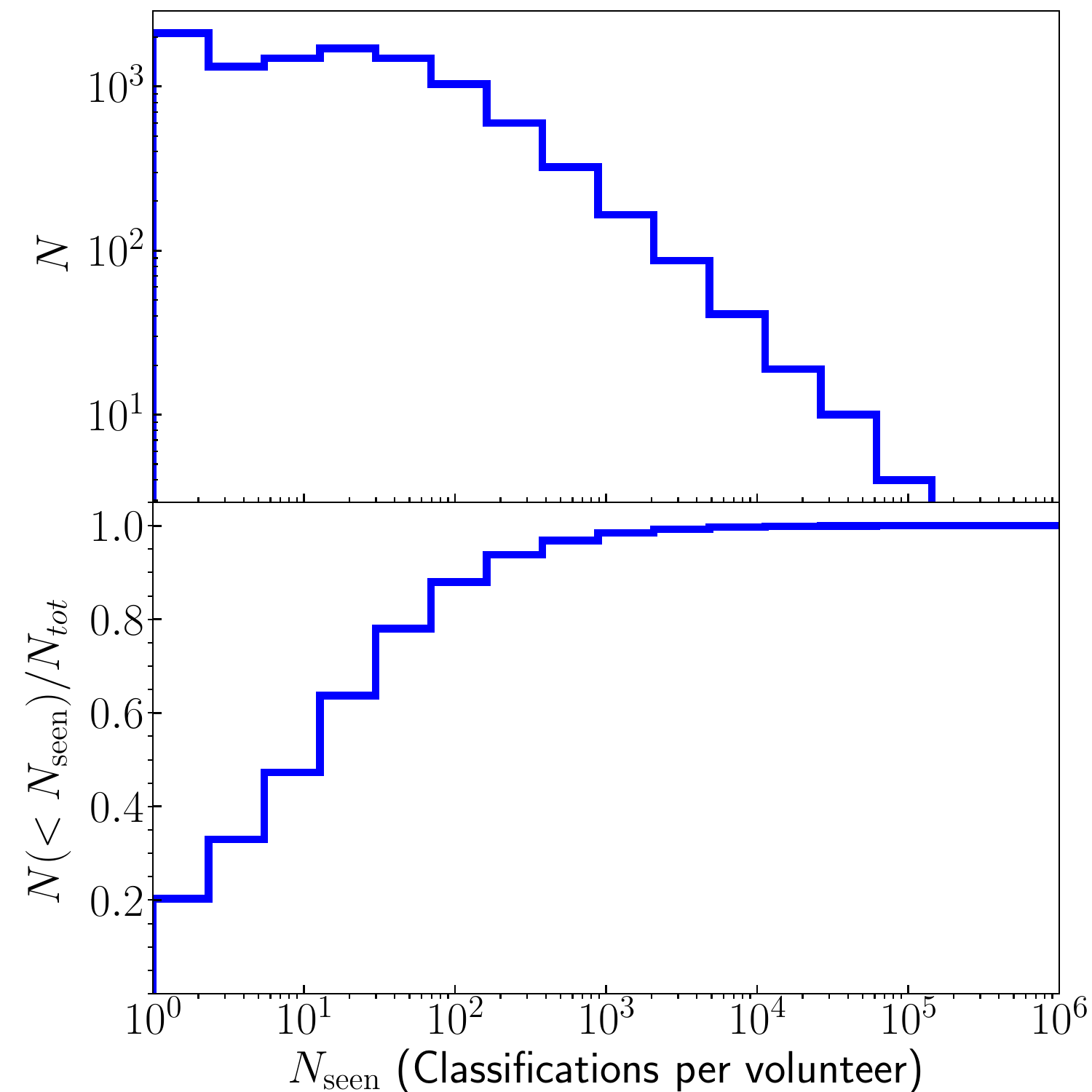}
\caption{
{\em Top:} Distribution in the number of classified subjects per volunteer. {\em Bottom:} Cumulative distribution.
\label{fig:nseen}
}
\end{figure}

The Space Warps~-~HSC crowdsourcing experiment was launched on April 27, 2018. It saw the participation of $\sim6000$ volunteers, who carried out $2.5$~million classifications over a period of two months. 
With the goal of assessing the degree of involvement of the volunteers, we show in \Fref{fig:nseen} the distribution in the number of classified subjects per user, $\nseen$. This is a declining function of $\nseen$, typical of crowdsourcing experiments: while most volunteers classified less than twenty subjects, nearly 20\% of them contributed each with at least a hundred classifications. It is thanks to these highly committed volunteers that the vast majority of the classifications needed for our analysis was gathered.

\begin{figure}
\includegraphics[width=\columnwidth]{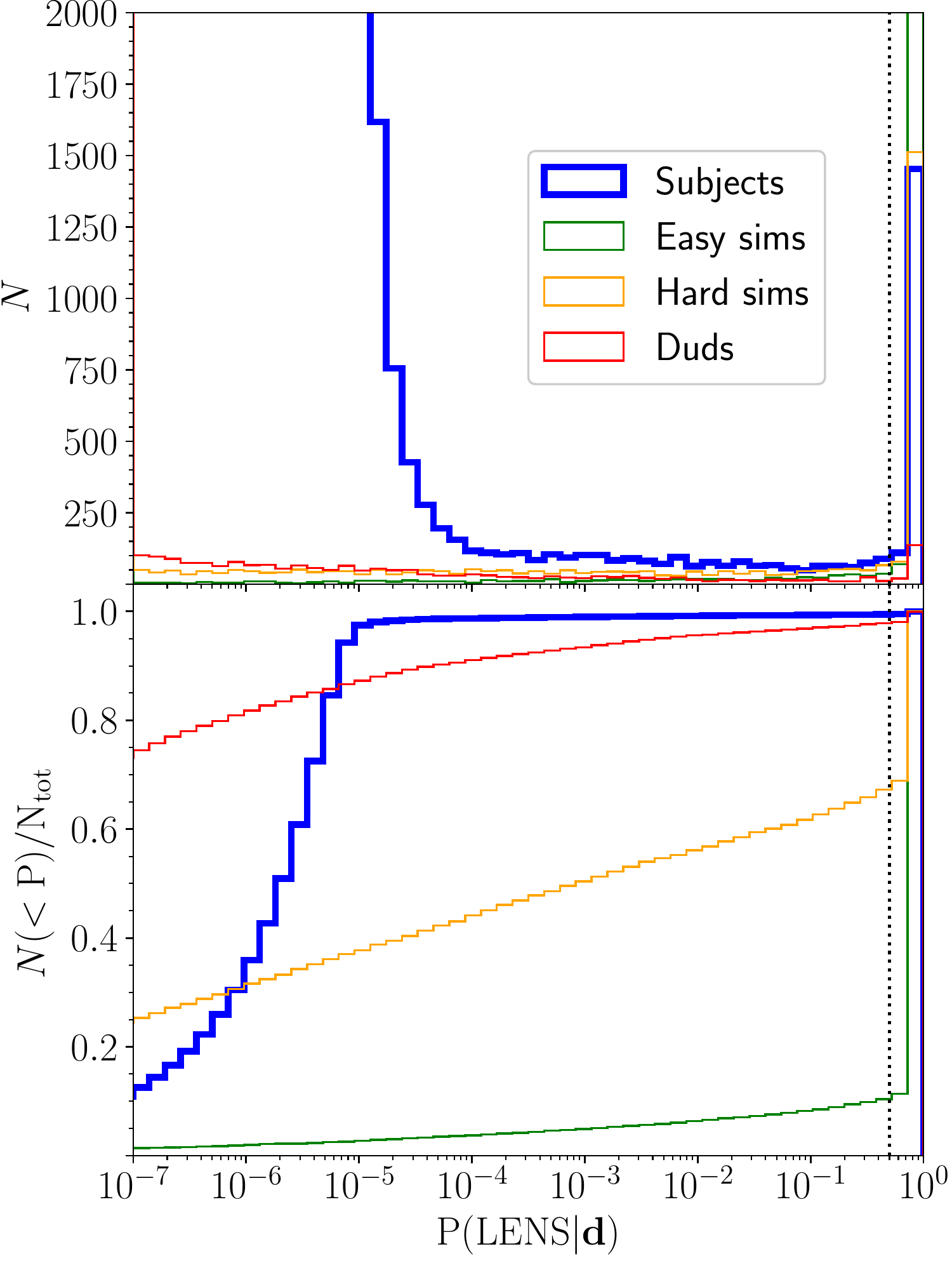}
\caption{
{\em Top:} Distribution in the posterior probability of subjects, duds, and sims being lenses given the classification data from the volunteers. {\em Bottom:} Cumulative distribution. The vertical dotted line marks the limit above which subjects are declared promising candidates and promoted to the expert visual inspection step.
\label{fig:plens}
}
\end{figure}

In \Fref{fig:plens}, we show the distribution of lens probabilities of the full sample of subjects (thick blue histograms), as quantified with the {\sc swap} software. 
This is a highly bimodal distribution: most of the subjects have very low lens probabilities, as expected, given the rare occurrence of the strong lensing phenomenon, but there is a high probability peak, corresponding to the objects identified as lenses by the volunteers.
We then made an arbitrary cut at $\pr(\rm{LENS}) = 0.5$: $\Nprob$ subjects with a lens probability that was larger than this value were declared `promising candidates' and promoted to the next step in our analysis, which consisted of visual inspection by the experts in strong lensing.
This further refinement of the lens candidate sample is described in detail in Subsection \ref{ssec:grading}.

In \Fref{fig:plens}, we also plot the distributions in lens probability of the three sets of training subjects: duds, easy sims, and hard sims.
These roughly follow  our expectations: most of the duds are correctly identified as such, with 90\% of the easy lenses having $\pr(\rm{LENS}) > 0.5$, while only a third of the hard lenses make it into our promising candidate cut. This validates our set of choices and criteria that went into compiling the easy and hard sims.
Of the 52 known lenses used for training, 49 were successfully classified as lenses (not shown in \Fref{fig:plens}). The three missed ones were all hard lenses.

With the goal of better quantifying the ability of the volunteers at finding lenses, we analysed the true positive rate (i.e. the fraction of lenses correctly classified as such over the total number of lenses inspected) on the simulated lenses as a function of lens properties.
In order for a human to recognise a gravitational lens, it is necessary for the lensed image to be detected with a sufficiently high signal-to-noise ratio (S/N) and for it to be spatially resolved, so that the image configuration typical of a strong lens can be observed. 
For this reason, we focused on two parameters: the S/N of the lensed images, summed within the footprint of the lensed source, and the angular size of the Einstein radius. We defined the lensed source footprint as the ensemble of pixels where the surface brightness of the lensed source is 3$\sigma$ above the sky background fluctuation level in the $g-$band.

\begin{figure*}
\begin{tabular}{cc}
\includegraphics[width=0.45\textwidth]{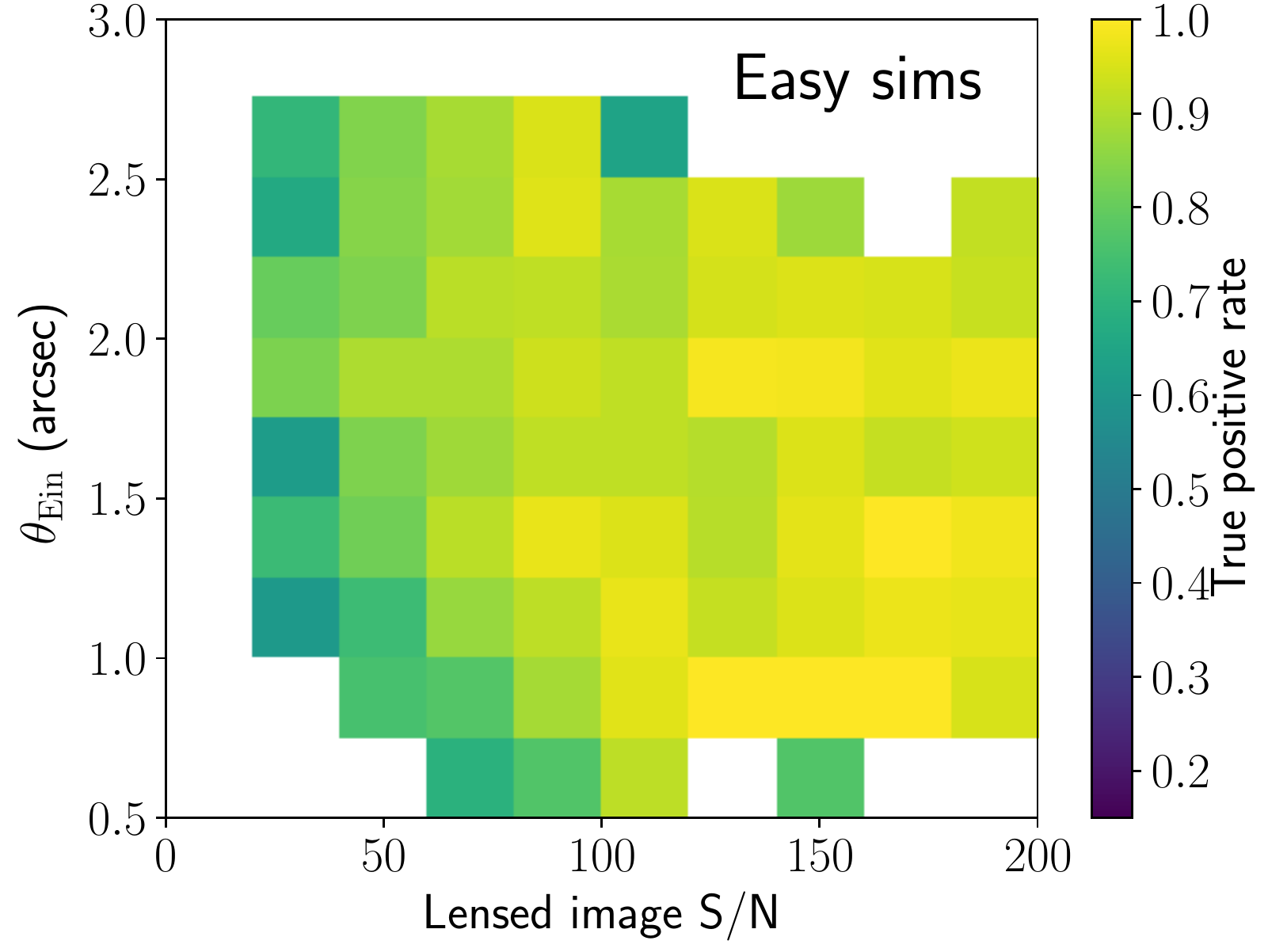} &
\includegraphics[width=0.45\textwidth]{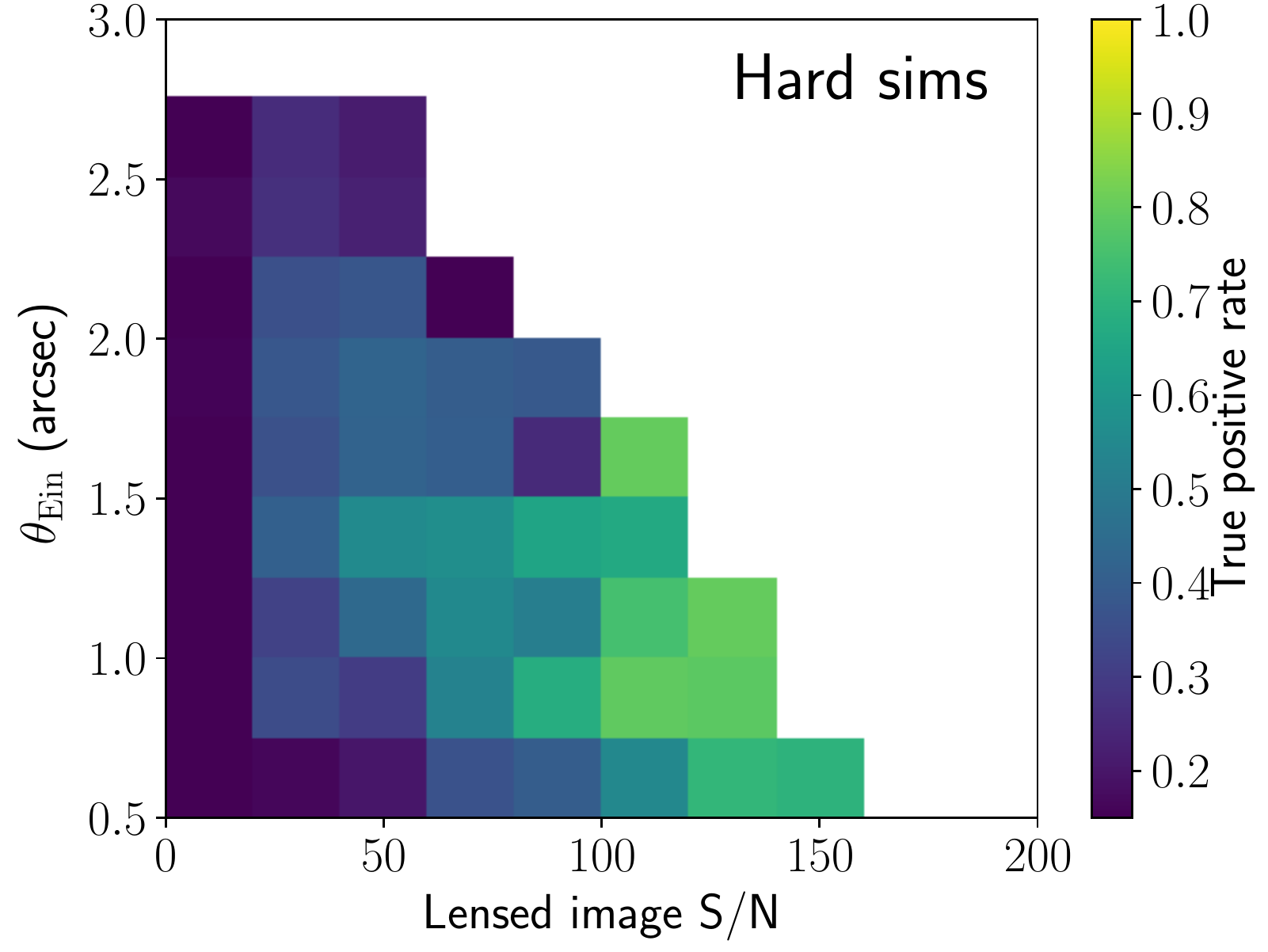} 
\end{tabular}
\caption{
True positive rate (i.e. fraction of lenses correctly classified as such) on the classification of simulated lenses as a function of lensed image S/N and lens Einstein radius for easy and hard lenses separately.
Parts of the parameter space occupied by only ten or fewer lenses are left white.
\label{fig:tcompl}
}
\end{figure*}

In \Fref{fig:tcompl}, we show the true positive rate on the simulated lenses, separated into easy and hard ones, as a function of lensed images S/N and $\tein$.
The volunteers were able to identify most of the easy lenses, largely regardless of $\tein$ and S/N. On the contrary, most of the hard lenses were missed, with the exception of systems with ${\rm S/N}>100$. 
The marked difference between the true positive rate on the easy and hard lenses of similar $\tein$ and S/N indicates once again that our criteria for the definition of the two categories are appropriate. As described in Subsection \ref{sssec:sims}, in order for a lens to be considered easy, the lensed image must have a high contrast with respect to the lens light and must consist of either two or more clearly visible images or one fairly extended arc.
It is reassuring that the volunteers achieved a high true positive rate on these systems, even for lenses with Einstein radius close to the resolution limit of $0.7''$.

At the same time, the extent of the low efficiency area among the hard lenses seen in \Fref{fig:tcompl} shows how the true positive rate depends nontrivially on a series of lens properties other than lensed image S/N and size of the Einstein radius, including the contrast between lens and source light and image configuration. This is true for volunteers and lensing experts alike, as we remark in Subsection \ref{ssec:diverse}, and in the absence of a simulation that realistically reproduces  the true distribution of strong lenses in all its detail makes it difficult to obtain an estimate of the completeness achieved by our crowdsourced search in HSC data. 

\subsection{Search with {\sc YattaLens}}\label{ssec:yatta}

\yatta\ is a lens finding algorithm developed by \citet{Son++18} consisting of two main steps.
In the first step, it runs {\sc SExtractor} \citep{B+A96} on foreground light-subtracted $g-$band images to look for tangentially elongated images (i.e. arcs) and possible counter-images with similar colours.
In the second step, it fits a lens model to the data and compared the goodness-of-fit with that obtained from two alternative non-lens models. In the cases when the lens model fits best, it keeps the system as a candidate strong lens.

We ran \yatta\ on the sample of $\sim300000$ galaxies obtained by applying the stellar mass and photometric redshift cuts described in Subsection \ref{ssec:sample} to the S16A internal data release catalogue of HSC.
More than 90\% of the subjects were discarded at the arc detection step. Of the remaining $\sim22000$, $\Nyatta$ were flagged as possible lens candidates by \yatta\ and the rest was discarded on the basis of the lens model not providing a good fit.
We then visually inspected the sample of lens candidates with the purpose of identifying non-lenses erroneously classified by \yatta\ that could be used for training purposes. The $\sim3800$ galaxies that made up the duds were drawn entirely from this sample.

\subsection{Lens candidate grading}\label{ssec:grading}

We merged the sample of lens candidates identified by the volunteers, $\Nprob$ subjects with $\pr(\rm{LENS}|\data) > 0.5$, with the \yatta\ sample, from which we removed the $\sim3800$ subjects used as duds. 

We also added to the sample $\Ntalk$ outstanding candidates flagged by volunteers on the `Talk' section of the Space Warps website, which we browsed on a roughly daily basis, quickly inspecting subjects that had recently been commented on (typically on the order of a few tens each day). This last subsample is by no means complete (we did not systematically inspect all subjects flagged by the volunteers) and it has a large overlap with the set of probable lenses produced by the classification algorithm. Nevertheless, we included it in order to make sure that potentially interesting candidates would not get lost. Although most of the candidates inspected in this way turned out not to be lenses, this step still proved to be useful because it enabled the discovery of a few lenses that would have otherwise been missed (as we show later in this paper).

We then visually inspected the resulting sample with the purpose of refining the candidate classification. Nine co-authors of this paper assigned to each candidate an integer score from 0 to 3 to indicate the likelihood of the subject being a strong lens.
We used the following scoring convention:
\begin{itemize}
\item ${\rm Score} = 3$: almost certainly a lens. A textbook example for which all characteristics of lensed images are verified: image configuration, consistency of colour and, in case of extended sources, surface brightness among all images. Additionally, the possibility of lensed features being the result of contamination can be ruled out with high confidence.
\item ${\rm Score} = 2$: probably a lens. All of the features match those expected for a strong lens, but the possibility that some of the features are due to contaminants cannot be ruled out.
\item ${\rm Score} = 1$: possibly a lens. Most of the features are consistent with those expected for a strong lens, but they may as well be explained by contaminants.
\item ${\rm Score} = 0$: almost certainly not a lens. Features are inconsistent with those expected for a strong lens.
\end{itemize}
Additionally, in order to ensure consistency in grading criteria across the whole sample and among different graders, we proposed the following algorithm for assigning scores.
\begin{enumerate}
\item Identify the images that could be lensed counterparts of each other
\item Depending on the image multiplicity and configuration, assign an initial score as follows:
  \begin{itemize}
  \item Einstein rings, sets of four or more images, sets consisting of at least one arc and a counter-image: 3 points.
  \item Sets of three or two images, single arcs: 2 points.
  \end{itemize}
\item If the lens is a clear group or cluster, add an extra point up to a maximum provisional score of 3.
\item Remove points based on how likely it is that the observed features are the result of contamination or image artefacts: if artefacts are present, then multiple images may not preserve surface brightness, may show mismatch of colours, or may have the wrong orientation or curvature around the lens galaxy.
\item Make sure that the final score is reasonable given the definitions outlined above.
\end{enumerate}
The rationale for the third point is to take into account the fact that a) groups and clusters are more likely to be lenses, due to their high mass concentration and b) often produce non-trivial image configurations which might be penalised during the fourth step.
Finally, we averaged the scores of all nine graders, and assigned a final grade as follows:
\begin{itemize}
\item Grade A: $\left<{\rm Score}\right> > 2.5$.
\item Grade B: $1.5 < \left<{\rm Score}\right> <= 2.5$.
\item Grade C: $0.5 < \left<{\rm Score}\right> <= 1.5$.
\item Grade 0: $\left<{\rm Score}\right> <= 0.5$.
\end{itemize}
We found $\NgradeA$ grade A lenses, $\NgradeB$ grade B, and $\NgradeC$ C lens candidates.
In \Tref{tab:summary} we provide a summary of the number of lens candidates of each grade found separately by Space Warps, both from the selection based on classification data and from the `Talk' section, and with \yatta.
\begin{table}
\caption{Number of lens candidates of grades of each grade found among subjects selected by the cut $\pr({\rm LENS}|\data) > 0.5$, from the `Talk' section of Space Warps, by \yatta\ and in the merged sample.
\label{tab:summary}
}
\begin{tabular}{l|ccccc}
\hline
\hline
Sample & A & B & C & 0 & \# Inspected\\
\hline
${\rm P}({\rm LENS}|\mathbf{d}) > 0.5$ & 14 & 118 & 465 & 980 & 1577\\
`Talk' & 11 & 84 & 121 & 48 & 264\\
{\sc YattaLens} & 6 & 67 & 233 & 6473 & 6779\\
\hline
Merged & 14 & 129 & 581 & 7152 & 7876\\
\end{tabular}
\end{table}

The first thing we can see from \Tref{tab:summary} is that, among the $\NgradeCabove$ lens candidates with grade C and above (sum of the first three columns in the bottom row), $597$ of them (82\%) are in the sample of subjects with $\pr({\rm LENS}|\data) > 0.5$. Only $11$ of the $\NgradeB$ grade B candidates and none of the grade A ones were missed by the analysis of volunteer classification data (i.e. have $\pr({\rm LENS}|\data) <= 0.5$).
In contrast, only about half of the grade A, B and C candidates were flagged as possible lenses by \yatta.
This clearly indicates that crowdsourcing returned a relatively more complete sample of candidates compared to \yatta. We discuss this and other differences in performance between the two methods in Subsection \ref{ssec:sw_vs_yatta}.

In \Tref{tab:gradeAB}, we list all the grade A and B candidates discovered.
The full list of lens candidates with grade C and better is provided online, in our database containing all lens candidates found or analysed by SuGOHI\footnote{\url{http://www-utap.phys.s.u-tokyo.ac.jp/~oguri/sugohi/} Candidates from this study are identified by the value `SuGOHI6' in the `Reference' field.
}, and as an ASCII table at the CDS\footnote{
CDS data can be retrieved via anonymous ftp to \url{cdsarc.u-strasbg.fr} (130.79.128.5)
or via \url{http://cdsweb.u-strasbg.fr/cgi-bin/qcat?J/A+A/}
}
.
This database was created by merging samples of lenses from our previous studies. Lens candidates that have been independently discovered as part of different lens searches can have different grades because of differences in the photometric data used for grading, including the size and colour scheme of the image cutouts or in the composition of the team who performed the visual inspection. In such cases, the higher grade was taken under the assumption that it was driven by a higher quality in the image cutout used for the inspection (in terms of the lensed features being more clearly visible). As a result, the number of grade A and B lens candidates from this study present in the SuGOHI database is slightly larger than the $\NgradeAB$ candidates listed in \Tref{tab:gradeAB}. This is due to an overlap between the lens candidates found in this work and those from our visual inspection of galaxy clusters by \citet[][, Paper V]{Jae++20b}, carried out in parallel.

\begin{figure}
\includegraphics[width=\columnwidth]{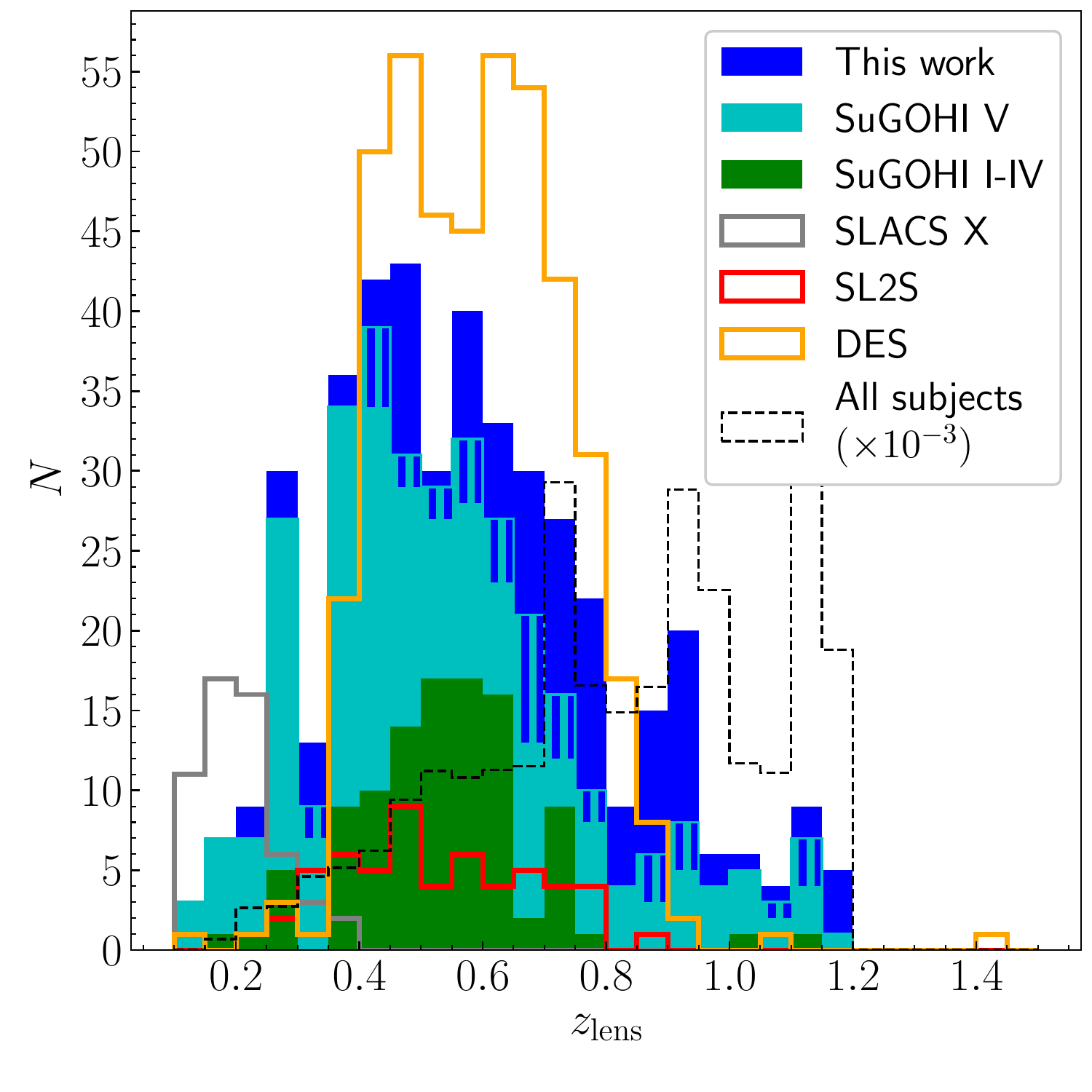}
\caption{
{\em Shaded region:} Distribution in lens photometric redshift of all grade A and B SuGOHI lenses. The blue portion of the histogram corresponds to lenses discovered in this study.
The green part indicates BOSS galaxy lenses discovered with \yatta, presented in Paper I, II, and IV\citep{Son++18, Won++20, Cha++20}.
The cyan part shows lenses discovered by means of visual inspection of galaxy clusters, as presented in Paper V. The striped regions indicate lenses discovered independently in this study and in the study of Paper V.
{\rm Grey solid lines:} distribution in lens spectroscopic redshift of lenses from the Sloan ACS Lens Survey \citep[SLACS][]{Aug++10}.
{\rm Red solid lines:} Distribution in lens spectroscopic redshift of lenses from the SL2S Survey \citep{Son++13a, Son++13b, Son++15}.
{\rm Orange solid lines:} Distribution in lens photometric redshift of likely lenses discovered in DES by \citet{Jac++19b}.
{\em Dashed lines:} distribution in photometric redshift of all subjects examined in this study, rescaled downwards by a factor of 1000.
\label{fig:photoz}
}
\end{figure}

One of the main goals of this experiment was to extend the sample of known lenses to higher lens redshifts. In \Fref{fig:photoz}, we plot a histogram with the distribution in photo-z of our grade A and B lens candidates, together with candidates from our previous searches and compared to the lens redshift distribution from other surveys.
The SuGOHI sample consists now of $\NsugohiAB$ highly probable (grade A or B) lens candidates. This is comparable to strong lens samples found in the Dark Energy Survey \citep[DES, where][ discovered 438 previously unknown lens candidates, one of which is also in our sample]{Jac++19b} and in the Kilo-Degree Survey \citep[KiDS:][ presented $\sim300$ new candidates, not shown in Figure \ref{fig:photoz} due to a lack of published redshifts, 15 of which are also in our sample]{Pet++19b}. 
Most notably, 41 of our lenses have photo-zs larger than $0.8$, which is more than any other survey.

Some caution is required when using photometric redshifts, however: 
the distribution in photo-z of all our subjects (shown as a dashed histogram in \Fref{fig:photoz}) shows unusual peaks around a few values, which appear to be reflected in the photo-z distribution of the lenses (dotted line). Given the large sky area covered by our sample, we would have expected a much smoother photo-z distribution. Therefore, we believe these peaks to be the result of systematic errors in the photo-z.
In order to obtain an estimate for the magnitude of such errors, we considered the subset of galaxies from our sample with spectroscopic redshift measurements from the literature \citep[see][ for details on the spectroscopic surveys overlapping with HSC]{Tan++18}. The distribution in spectroscopic redshift of galaxies with photo-z larger than $1.0$ has a median value of $1.05$ and a tail that extends towards low redshifts. The tenth percentile of this distribution is at a spectroscopic redshift of $0.65$. Assuming that the distribution in spectroscopic redshift of the lens sample follows a similar distribution, we can use this as a lower limit to the true redshift of our $z_{\mathrm{phot}} > 1.0$ candidates.
Since we are not using photo-z information to perform any physical measurement, we defer any further investigation of photo-z systematics to future studies.

\subsection{A diverse population of lenses}\label{ssec:diverse}

\begin{figure}
\includegraphics[width=\columnwidth]{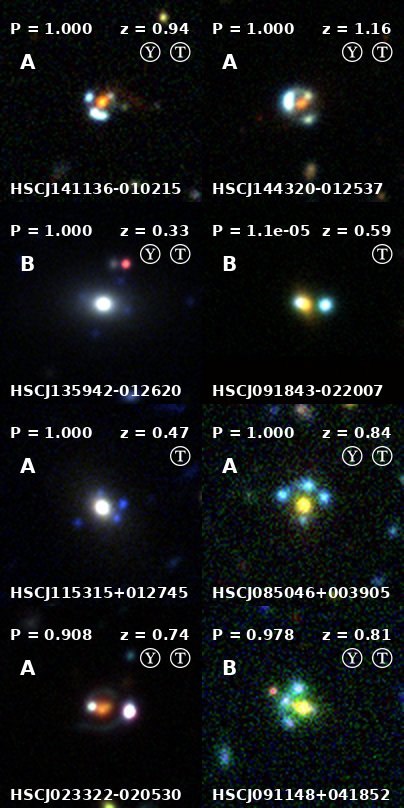}
\caption{
Set of eight lenses associated with a compact lensed source. In each panel, we show the probability of the subject of being a lens, according to the volunteer classification data (top left), the lens galaxy photo-z (top right), the final grade after our inspection (left). The circled 'Y' and 'T' on the right, if present, indicate that the candidate was discovered by \yatta\ and noted by us in the `Talk' section of the Space Warps website, respectively.
\label{fig:compact}
}
\end{figure}

In the rest of this section, we highlight a selected sample of lens candidates that we find interesting, grouped by type.
We begin by showing in \Fref{fig:compact} a set of eight lenses with a compact background source, that is, lenses with images that are visually indistinguishable from a point source. Compact strongly lensed sources are interesting because they could be associated with active galactic nuclei or, alternatively, they could be used to measure the sizes of galaxies that would be difficult to resolve otherwise \citep[see e.g.][]{Mor++17, Jae++20a}.
The two lenses in the top row of \Fref{fig:compact} were also featured in the fourth paper of our series, dedicated to a search for strongly lensed quasars \citep{Cha++20}. 
All lenses shown in \Fref{fig:compact} were classified as such by the volunteers (the value of $\pr({\rm LENS}|\mathbf{d})$ is shown in the top left corner of each image), with the exception of HSCJ091843.38$-$022007.3.
We were able to include it in our sample thanks to a single volunteer who flagged it in the `Talk' section. 

\begin{figure}
\includegraphics[width=\columnwidth]{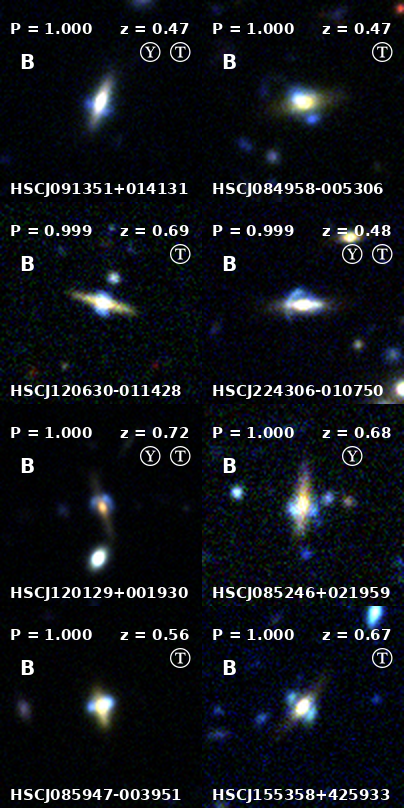}
\caption{
Set of eight disk galaxy lens candidates discovered in our sample.
\label{fig:disks}
}
\end{figure}

Eight of the grade B candidates, shown in \Fref{fig:disks}, and more among the grade C ones, have a disk galaxy as lens.
Disk galaxies represent a minority among all lenses and most of the previously studied ones are at a lower redshift compared to the average of our sample \citep[but see][ for an exception]{Suy++12}. The largest sample of disk lenses studied so far is the Sloan WFC Edge-on Late-type Lens Survey \citep[SWELLS][]{Tre++11}, which consists of 19 lenses at $z<0.3$. Our newly discovered disk lenses extend this family of objects to higher redshift, and, with appropriate follow-up observations, could be used to study the evolution in the mass structure of disks\footnote{Although the mass within the Einstein radius of these systems is likely dominated by the bulge}.

\begin{figure}
\includegraphics[width=\columnwidth]{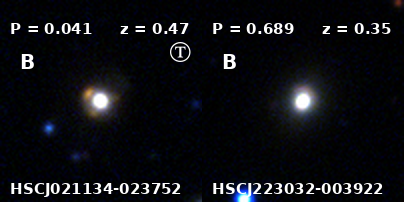}
\caption{
Two lens candidates with strongly lensed red sources.
\label{fig:red}
}
\end{figure}

Most of the lensed sources in our sample have blue colours. This is related to the fact that the typical source redshift is in the range $1 < z < 3$ \citep[see e.g.][]{Son++19b}, close to the peak of cosmic star formation activity.
 Nevertheless, we were able to discover a limited number of lenses with a red background source. Two of them were classified as grade B candidates and are shown in \Fref{fig:red}. 
The object on the left has a standard fold configuration and was conservatively given a grade B only because one of the images is barely detected in the data. However, it was not classified as lens by the volunteers (although the final value of $\pr({\rm LE NS}|\mathbf{d})$ is higher than the prior probability). 
Our training sample consisted almost exclusively of blue sources: it is then possible that the volunteers were not ready to recognise such an unusual lens \citep[although past crowdsourcing experiments proved otherwise,][]{Gea++15}.
We included it in the sample after it was flagged by one volunteer in the `Talk' section.
Both lenses in \Fref{fig:red} were missed by \yatta, as it was set up to discard red arcs in order to eliminate contaminants in the form of neighbouring tangentially aligned galaxies.

Finally, we point out how roughly half of our grade C lens candidates consist of systems with either a single visible arc, the counter-image of which, if present, is very close to the centre and not detected, or a double in a highly asymmetric configuration.
It is very difficult to determine whether such candidates are lenses or not using photometric data alone but given their abundance (a few hundred in the whole sample), they constitute a very interesting category of lenses: even if only a fraction of them turned out to be real lenses, they would end up dominating the lens population.
This is not surprising but, rather, it is to be expected from the simple geometrical arguments that we made when describing our procedure for simulating lenses, in Subsection \ref{sssec:sims}: the area in the source plane that gets mapped into highly asymmetric image configurations is larger than the area corresponding to configurations that are close to symmetric\footnote{The picture is complicated by magnification effects: more symmetric configurations correspond in general to a higher magnification, allowing to detect fainter sources in a flux-limited survey. The net effect depends on the slope of the source luminosity function.}.

\begin{figure*}
\begin{tabular}{cc}
\includegraphics[width=0.47\textwidth]{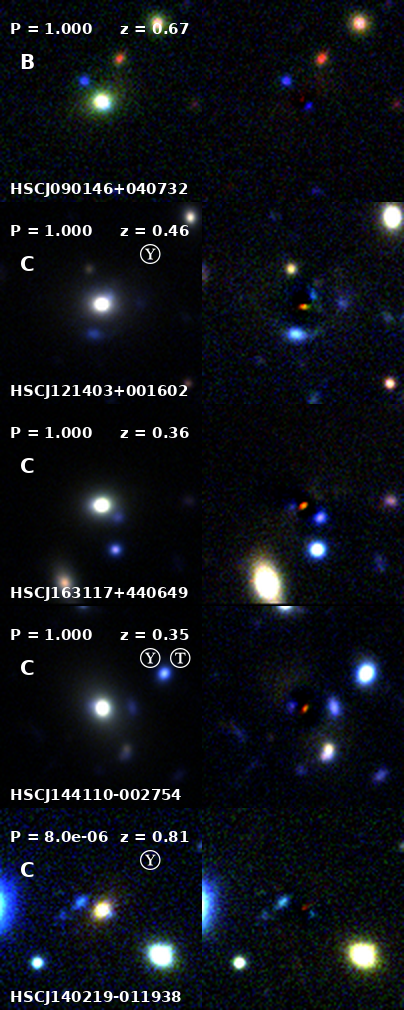} &
\includegraphics[width=0.47\textwidth]{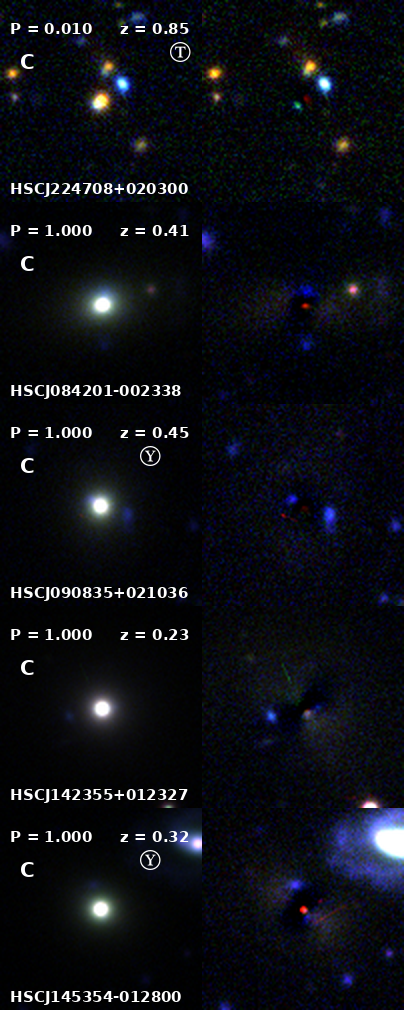}
\end{tabular}
\caption{Selected sample of ten lens candidates with a highly asymmetric image configuration. In each row, the left panel shows the original image, while the right one shows the foreground-subtracted one.}
\label{fig:asymm}
\end{figure*}

In \Fref{fig:asymm}, we show a collection of some of the best examples of highly asymmetric doubles that we were able to discover.
The figure highlights the importance of the foreground light subtraction step which, although far from perfect (large negative residuals are typically left in the centre of the image), helps greatly in the detection of faint counter-images close to the centre.
Such asymmetric systems are interesting because they allow constraints to be put on the mass in the very inner regions of a lens, which is dominated by the stars and with a possible contribution from the central supermassive black hole, even in cases when a counter-image is not detected \citep{SLC18, Smi++20}.
Incidentally, \Fref{fig:asymm} also illustrates the difficulty in assigning consistent grades to large samples of lens candidates: the object in the top left was given a grade B, in accordance to the criteria discussed in Subsection \ref{ssec:grading}, while all the other ones were assigned a grade C despite having a very similar image configuration.
In our past searches, we used to collectively re-discuss lens candidate grades on a one-by-one basis after a first round of inspection. This, however, was not feasible in the present study due to the large data volume.

\section{Discussion}\label{sect:discuss}

\subsection{Performance of different lens finding methods}\label{ssec:sw_vs_yatta}

The two most important quantities that define the performance of a lens-finding method are completeness and purity. The former is the fraction of strong lenses among the ones present in the surveyed sample that are recovered by the method, while the latter is the fraction of objects among the ones labelled as strong lenses that are indeed lenses.
Unfortunately, it is very difficult to determine either of them in an absolute sense: it would require us to apply our lens-finding methods to a large complete sample of real lenses and to a large sample of galaxies representative of our survey the nature of which is known exactly.
We can, however, evaluate the relative performance between our two methods, namely, crowdsourcing and \yatta.
The data reported in \Tref{tab:summary} shows clearly how the former outperformed the latter both in terms of completeness, with roughly twice the number of lens candidates with grade C or higher, and purity, with 40\% of the inspected candidates having grade C or higher, against only 5\% for \yatta.
The comparison is not entirely fair: first of all, \yatta\, correctly identified most of the 52 known lenses used in the training sample, which have been excluded from the summary of \Tref{tab:summary} and would have otherwise increased the completeness of \yatta\ \citep[many of these lenses belong to the sample of lenses discovered with \yatta\ by][]{Son++18}. Secondly, $3800$ duds initially found by \yatta\ were removed from the subject sample and only shown to the volunteers as training images, making their classification job slightly easier. However, given the relatively good performance of the volunteers on the training subjects, with only $\sim10\%$ of the duds being classified as lenses (see \Fref{fig:plens}), the purity of the sample produced by the volunteers would only have changed by a small margin by including the duds.

Most of the lens candidates missed by \yatta\ were discarded at the arc detection step for various reasons: either their lensed images are point-like (as two of the candidates shown in \Fref{fig:compact}) or too red (as in the two cases shown in \Fref{fig:red}), or consist of arcs that are considered too faint or too far from the lens galaxy by \yatta.
In principle, we could adjust the settings of \yatta\ to be able to detect such lenses in future searches, although most likely entailing a penalty in terms of purity.

In the previous section, we also reported the discovery of a number of highly probable lenses that were flagged by volunteers in the `Talk' section of Space Warps, including some that were missed by both the volunteer classification data and \yatta.
Based on the numbers reported in \Tref{tab:summary}, with as many as 11 grade A and 86 grade B lens candidates found among 264 inspected candidates, we could be inclined to conclude that the `Talk' section provides much purer samples compared to the analysis of classification data. However, those numbers are misleading: 264 were the candidates that were deemed sufficiently interesting to be included in the final grading step, and were selected after sorting through thousands of subjects flagged by volunteers. The effective purity of this lens search method is therefore much lower than suggested by \Tref{tab:summary}.

\subsection{Comparison with the \citet{Met++19} lens-finding challenge}

\citet{Met++19} carried out a lens finding challenge, in which $100000$ simulated images of lenses and non-lenses were classified with a variety of lens finding methods over the course of 48 hours. Among the lens finders who took part in the challenge, there were several machine learning-based methods, a visual inspection effort led by strong lensing experts, and a simplified version of \yatta, dubbed {\sc YattaLens Lite}, limited to the arc-finding step and stripped of the modelling part to meet the time constraints of the challenge. The methods characterised by the best performance were based on machine learning, some of which achieved better results than visual inspection.
{\sc YattaLens Lite} achieved a false positive rate of $\sim10\%$ and a true positive rate of $75\%$, while the performance of visual inspection was marginally better.

The performance of \yatta\ on the real data used in this work is different from that of {\sc YattaLens Lite} on the lens finding challenge. We achieved a false positive rate of $\approx 2\%$ (given by the number of grade 0 candidates classified as lenses by \yatta, $6470$, among the $300000$ scanned subjects). This lower value can be explained partly by the presence of the modelling step, which was skipped in the lens-finding challenge and which typically brings an improvement in purity of a factor of three, and partly by the different composition of the non-lens sample of the challenge as compared to the sample of real galaxies.
The true positive rate is also lower in this experiment: although we do not know the total number of lenses present among all scanned subjects, we can obtain an upper limit on the true positive rate by dividing the number of grade A and B candidates recovered by \yatta\ ($6+68=74$) by the total number of grade A and B candidates found ($14+130=144$), roughly $50\%;$ this fraction increases if we also consider the $52$ real lenses used as training subjects, but it is still below the 75\% true positive rate scored in the lens finding challenge.
As for the false positive rate, the true positive rate is also sensitive to the details of the distribution of lens properties in the sample: for example, we suspect that the lens finding challenge had a higher fraction of lenses that would be classified as `easy' according to our definition, boosting the true positive rate.
Indeed, the lens sample of \citet{Met++19} is known for having an unusually large number of lenses with a large Einstein radius ($4''$ or more), as also noted by \citet{Dav++19}.

The main lesson from this comparison is that while lens-finding challenges carried out on simulated data can be very useful tests of lens finding methods, results can vary a lot depending on the details of the test samples used.
Therefore, tests on real data are essential to accurately assess the performance of a given lens finding method. These are not a viable option at the moment due to the relatively low number of known lenses but it might become feasible in the future.

\subsection{Lens-finding efficiency dependence on image depth}\label{ssec:depth}

\begin{figure}
\includegraphics[width=\columnwidth]{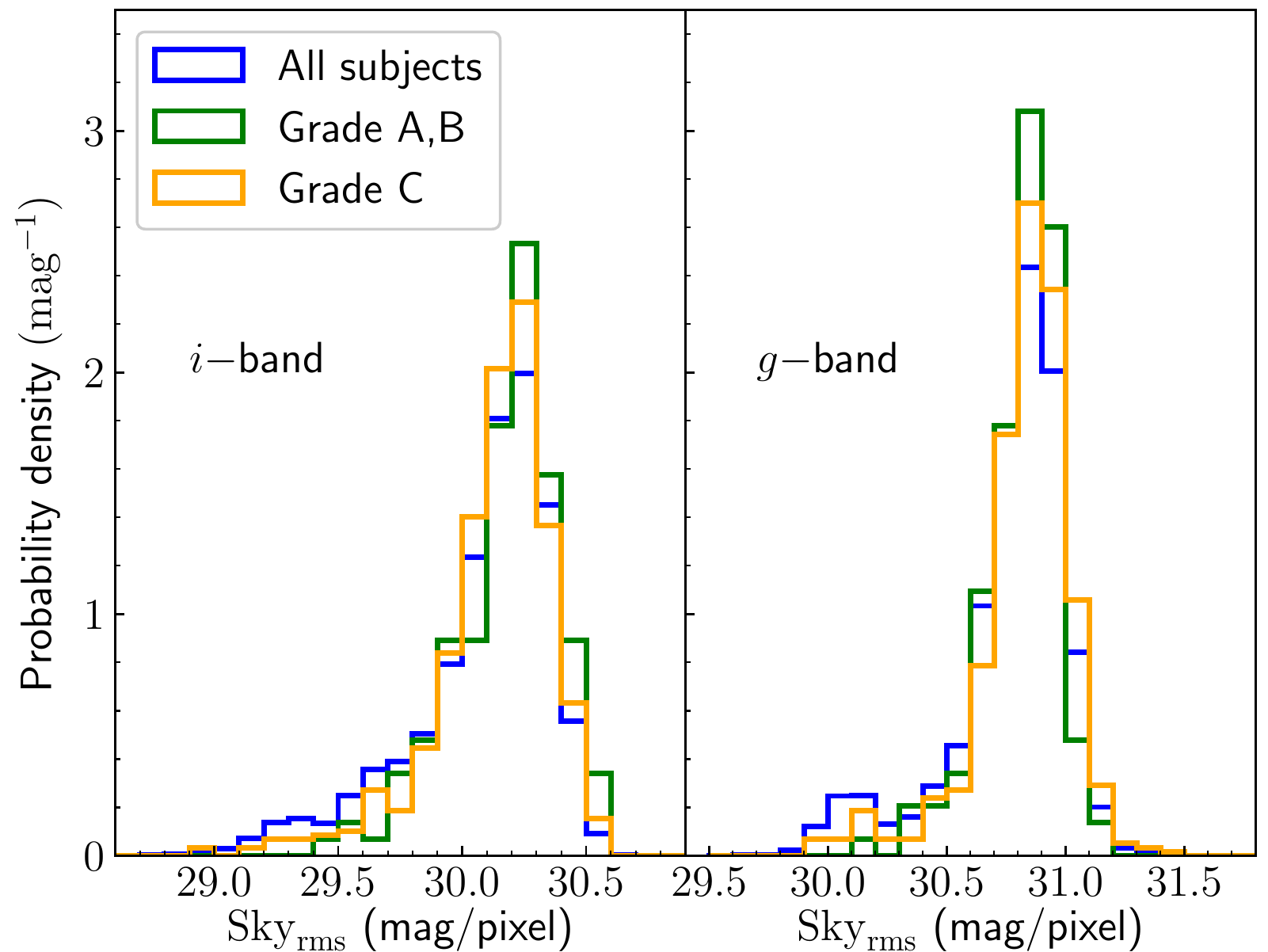}
\caption{Distribution in sky background rms (in AB magnitudes per pixel) of all subjects, grade A and B candidates combined, and grade C candidates, in $i-$ (left) and $g-$band (right).
\label{fig:skybkg}
}
\end{figure}

One of the most important aspects of photometric data for lens finding purposes is image depth: in principle, deeper data should allow the detection of fainter background sources, and therefore more lenses.
The data used for our study, taken from the S17A internal data release of HSC, span a wide range in depth: the number of individual exposures that make up the coadded images used for our analysis goes from a minimum of one to the survey standard value of six in $i-$band, and even more in regions where multiple pointings overlap.
We can then check whether the number density of lens candidates correlates with image depth. In \Fref{fig:skybkg} we plot the distribution in $i-$ and $g-$band sky background fluctuation of all subjects, of grade A and B lens candidates combined, and of grade C ones.
By looking at the $i-$band distribution (left panel), we can see how the distribution of grade A and B candidates is shifted towards lower levels of background noise compared to the distribution of all subjects. A Kolmogorov-Smirnov test reveals a p-value of $8.8\times10^{-4}$, hence a low probability that the two samples (all subjects and grade A and B candidates) are drawn from the same distribution.
While the $i-$band data confirms the idea that deeper data leads to a higher number of detected lenses, the $g-$band distribution appears to tell a different story: there is no obvious difference between the distribution in the background fluctuation of lens candidates and all subjects, with the Kolmogorov-Smirnov test giving a p-value of $0.16$.
Given that the $g-$band is, for the vast majority of our candidates, the one with the highest contrast between lens and source, we would have expected an even higher difference between the two distributions.
This result instead suggests that $g-$band depth is probably not the limiting factor in our lens finding campaign, but $i-$band depth is more important. This could be related to the foreground light subtraction step, for which we rely on the $i-$band image to obtain a model for the surface brightness profile of the lens. 

\subsection{Comparison with past crowdsourcing experiments}

This was the third crowd-sourced lens finding experiment carried out by Space Warps following the search based on CFHT-LS data \citep{Mar++16, Mor++16} and on the VISTA-CFHT Stripe 82 survey \citep{Gea++15}.
\citet{Mor++16} were able to find 91 probable lenses, 29 of which were previously unknown, in 160 square degrees of data. The number density of grade A and B lenses found in the present search is similar, though it is hard to make a quantitative comparison, due to different definitions of what constitutes a probable lens candidate and to the fact that we excluded known lenses from our subject sample from the start.
An essential difference between the present study and the CFHT-LS Space Warps campaign is that this was a targeted search: we looked for lenses only among a set of galaxies in a given redshift and stellar mass range. The search carried out by \citet{Mor++16} instead consisted of two stages: a blind search over tiles of the whole survey area, followed by a re-inspection of the most promising candidates.
Compared to the \citet{Mor++16} study, we found a much larger fraction of `undecided' subjects, for which their probability of being a lens did not converge neither to a value close to unity nor to very low values, and they were retired from the sample only after reaching the maximum allowed number of classifications. 
We think this is a consequence of the fact that volunteers have been able to detect fainter lens-like features that are intrinsically more difficult to classify as compared to the CFHT-LS campaign.
There are a few reasons for this. First of all, the chances of finding faint arcs are higher in a targeted search, when the attention is focused on a well-defined object, as opposed to a blind search. Secondly, HSC data is deeper than CFHT-LS, increasing the number density of fainter features in the vicinity of a foreground galaxy.  Thirdly, the presence of foreground-subtracted images in our experiment allows the identification of lenses with a lower contrast between source and lens light. These factors lead to a higher fraction of ambiguous lens candidates in our sample.

\section{Conclusions}\label{sect:concl}

We carried out a crowdsourced lens search based on over 442 square degrees of data from the HSC survey.
The search was carried out on a sample of $\sim300000$ galaxies with photometric redshift between $0.2$ and $1.2$ and a stellar mass larger than $10^{11.2}M_\odot$.
Almost $6000$ citizen volunteers participated in the crowdsourcing experiment, named Space~Warps~-~HSC. We collected $\sim2.5$~million classifications, which we then analysed with an algorithm developed in past editions of Space Warps.
In parallel, we searched for lenses in the same sample of galaxies using the automated lens finding method \yatta.
From the two searches combined, we found $\NgradeAB$ highly probable (grade A or B) new lens candidates, in an area that already included $\Nknown$ known lenses.
Compared to \yatta, crowdsourcing was by far the most successful lens finding method, both in terms of completeness and purity. We found lenses of a variety of kinds, including lenses with a compact source, with a red source, group-scale lenses, and lens candidates with highly asymmetric configurations.
From an analysis of the performance of crowdsourcing on simulated lens images, we determined that volunteers are able to correctly classify most lenses, including those with Einstein radius comparable to the resolution limit, as long as multiple images are clearly detected.

In the coming years, the volume of data available for lens finding purposes will increase greatly, as the Euclid space telescope\footnote{\url{https://euclid-ec.org/}} and the Large Synoptic Survey Telescope\footnote{\url{https://lsst.org}} (LSST) are each planned to cover areas of the sky that are more than a factor of 30 larger than that scanned in our study.
Scaling up Space Warps to such data volumes will be challenging: a much larger number of volunteers and a higher number of classification per volunteer will be needed. 
We can aim to improve the efficiency of our search by modifying the definition of the parent sample of subjects passed on to the volunteers. For instance, machine learning-based methods could potentially be used to pre-select large samples of possible lenses that are then refined in a visual inspection step via crowdsourcing.
Nevertheless, our experiment shows how crowdsourcing is a very powerful tool for finding lenses and delivering samples of lens candidates with relatively high purity and completeness, and so, we expect it to play a major role in lens finding in the 2020s.
\yatta\ can produce samples of roughly one grade-C lens candidate or better per square degree in HSC-like data and it is, therefore, confirmed to be a useful and efficient tool for finding galaxy-scale lenses in a semi-automated way provided that high completeness is not a critical requirement.

\begin{acknowledgements}

We thank all of the volunteers who participated in Space Warps~-~HSC, whose Zooniverse usernames are listed at the end of \url{https://www.zooniverse.org/projects/aprajita/space-warps-hsc/about/team}, with particular thanks to Ivan~A.~Terentev for flagging the lens HSCJ021134.26$-$023752.4 and many other good candidates.
We thank all of the Science Friday Team (in particular Ariel Zych, Christopher Intagliata, Brandon Echter and Ira Flatow) for featuring the SW-HSC project in two broadcasts, that greatly enhanced and promoted the project to a wider community. 
We also thank Michael Laraia, Hugh Dickinson and Lucy Fortson (University of Minnesota) for use of their implementation of the Space Warps Analysis Pipeline for the PANOPTES Zooniverse platform and integration with CAESAR functionality.

AS acknowledges funding from the European Union's Horizon 2020 research and innovation programme under grant agreement No 792916, as well as a KAKENHI Grant from the Japan Society for the Promotion of Science (JSPS), MEXT, Number JP17K14250.
ATJ is supported by JSPS KAKENHI Grant number JP17H02868.
JHHC acknowledges support from the Swiss National Science Foundation
(SNSF).
This work was supported by World Premier International Research Center Initiative (WPI Initiative), MEXT, Japan. 

The Hyper Suprime-Cam (HSC) collaboration includes the astronomical communities of Japan and Taiwan, and Princeton University.  The HSC instrumentation and software were developed by the National Astronomical Observatory of Japan (NAOJ), the Kavli Institute for the Physics and Mathematics of the Universe (Kavli IPMU), the University of Tokyo, the High Energy Accelerator Research Organization (KEK), the Academia Sinica Institute for Astronomy and Astrophysics in Taiwan (ASIAA), and Princeton University.  Funding was contributed by the FIRST programme from Japanese Cabinet Office, the Ministry of Education, Culture, Sports, Science, and Technology (MEXT), the Japan Society for the Promotion of Science (JSPS),  Japan Science and Technology Agency  (JST),  the Toray Science  Foundation, NAOJ, Kavli IPMU, KEK, ASIAA,  and Princeton University.

Funding for SDSS-III has been provided by the Alfred P. Sloan Foundation, the Participating Institutions, the National Science Foundation, and the U.S. Department of Energy Office of Science. The SDSS-III web site is http://www.sdss3.org/.

SDSS-III is managed by the Astrophysical Research Consortium for the Participating Institutions of the SDSS-III Collaboration including the University of Arizona, the Brazilian Participation Group, Brookhaven National Laboratory, Carnegie Mellon University, University of Florida, the French Participation Group, the German Participation Group, Harvard University, the Instituto de Astrofisica de Canarias, the Michigan State/Notre Dame/JINA Participation Group, Johns Hopkins University, Lawrence Berkeley National Laboratory, Max Planck Institute for Astrophysics, Max Planck Institute for Extraterrestrial Physics, New Mexico State University, New York University, Ohio State University, Pennsylvania State University, University of Portsmouth, Princeton University, the Spanish Participation Group, University of Tokyo, University of Utah, Vanderbilt University, University of Virginia, University of Washington, and Yale University.

\end{acknowledgements}

\bibliographystyle{aa}
\bibliography{references}

\begin{thebibliography}{63}
\expandafter\ifx\csname natexlab\endcsname\relax\def\natexlab#1{#1}\fi

\bibitem[{{Aihara} {et~al.}(2018{\natexlab{a}}){Aihara}, {Arimoto},
  {Armstrong}, {Arnouts}, {Bahcall}, {Bickerton}, {Bosch}, {Bundy}, {Capak},
  {Chan}, {Chiba}, {Coupon}, {Egami}, {Enoki}, {Finet}, {Fujimori}, {Fujimoto},
  {Furusawa}, {Furusawa}, {Goto}, {Goulding}, {Greco}, {Greene}, {Gunn},
  {Hamana}, {Harikane}, {Hashimoto}, {Hattori}, {Hayashi}, {Hayashi},
  {He{\l}miniak}, {Higuchi}, {Hikage}, {Ho}, {Hsieh}, {Huang}, {Huang},
  {Ikeda}, {Imanishi}, {Inoue}, {Iwasawa}, {Iwata}, {Jaelani}, {Jian},
  {Kamata}, {Karoji}, {Kashikawa}, {Katayama}, {Kawanomoto}, {Kayo}, {Koda},
  {Koike}, {Kojima}, {Komiyama}, {Konno}, {Koshida}, {Koyama}, {Kusakabe},
  {Leauthaud}, {Lee}, {Lin}, {Lin}, {Lupton}, {Mand elbaum}, {Matsuoka},
  {Medezinski}, {Mineo}, {Miyama}, {Miyatake}, {Miyazaki}, {Momose}, {More},
  {More}, {Moritani}, {Moriya}, {Morokuma}, {Mukae}, {Murata}, {Murayama},
  {Nagao}, {Nakata}, {Niida}, {Niikura}, {Nishizawa}, {Obuchi}, {Oguri},
  {Oishi}, {Okabe}, {Okamoto}, {Okura}, {Ono}, {Onodera}, {Onoue}, {Osato},
  {Ouchi}, {Price}, {Pyo}, {Sako}, {Sawicki}, {Shibuya}, {Shimasaku},
  {Shimono}, {Shirasaki}, {Silverman}, {Simet}, {Speagle}, {Spergel},
  {Strauss}, {Sugahara}, {Sugiyama}, {Suto}, {Suyu}, {Suzuki}, {Tait},
  {Takada}, {Takata}, {Tamura}, {Tanaka}, {Tanaka}, {Tanaka}, {Tanaka},
  {Terai}, {Terashima}, {Toba}, {Tominaga}, {Toshikawa}, {Turner}, {Uchida},
  {Uchiyama}, {Umetsu}, {Uraguchi}, {Urata}, {Usuda}, {Utsumi}, {Wang}, {Wang},
  {Wong}, {Yabe}, {Yamada}, {Yamanoi}, {Yasuda}, {Yeh}, {Yonehara}, \&
  {Yuma}}]{Aih++18}
{Aihara}, H., {Arimoto}, N., {Armstrong}, R., {et~al.} 2018{\natexlab{a}},
  \pasj, 70, S4

\bibitem[{{Aihara} {et~al.}(2018{\natexlab{b}}){Aihara}, {Armstrong},
  {Bickerton}, {Bosch}, {Coupon}, {Furusawa}, {Hayashi}, {Ikeda}, {Kamata},
  {Karoji}, {Kawanomoto}, {Koike}, {Komiyama}, {Lang}, {Lupton}, {Mineo},
  {Miyatake}, {Miyazaki}, {Morokuma}, {Obuchi}, {Oishi}, {Okura}, {Price},
  {Takata}, {Tanaka}, {Tanaka}, {Tanaka}, {Uchida}, {Uraguchi}, {Utsumi},
  {Wang}, {Yamada}, {Yamanoi}, {Yasuda}, {Arimoto}, {Chiba}, {Finet},
  {Fujimori}, {Fujimoto}, {Furusawa}, {Goto}, {Goulding}, {Gunn}, {Harikane},
  {Hattori}, {Hayashi}, {He{\l}miniak}, {Higuchi}, {Hikage}, {Ho}, {Hsieh},
  {Huang}, {Huang}, {Imanishi}, {Iwata}, {Jaelani}, {Jian}, {Kashikawa},
  {Katayama}, {Kojima}, {Konno}, {Koshida}, {Kusakabe}, {Leauthaud}, {Lee},
  {Lin}, {Lin}, {Mandelbaum}, {Matsuoka}, {Medezinski}, {Miyama}, {Momose},
  {More}, {More}, {Mukae}, {Murata}, {Murayama}, {Nagao}, {Nakata}, {Niida},
  {Niikura}, {Nishizawa}, {Oguri}, {Okabe}, {Ono}, {Onodera}, {Onoue}, {Ouchi},
  {Pyo}, {Shibuya}, {Shimasaku}, {Simet}, {Speagle}, {Spergel}, {Strauss},
  {Sugahara}, {Sugiyama}, {Suto}, {Suzuki}, {Tait}, {Takada}, {Terai}, {Toba},
  {Turner}, {Uchiyama}, {Umetsu}, {Urata}, {Usuda}, {Yeh}, \&
  {Yuma}}]{Aih++18b}
{Aihara}, H., {Armstrong}, R., {Bickerton}, S., {et~al.} 2018{\natexlab{b}},
  \pasj, 70, S8

\bibitem[{{Auger} {et~al.}(2010){Auger}, {Treu}, {Gavazzi}, {Bolton},
  {Koopmans}, \& {Marshall}}]{Aug++10}
{Auger}, M.~W., {Treu}, T., {Gavazzi}, R., {et~al.} 2010, \apjl, 721, L163

\bibitem[{{Axelrod} {et~al.}(2010){Axelrod}, {Kantor}, {Lupton}, \&
  {Pierfederici}}]{Axe++10}
{Axelrod}, T., {Kantor}, J., {Lupton}, R.~H., \& {Pierfederici}, F. 2010, in
  \procspie, Vol. 7740, Software and Cyberinfrastructure for Astronomy, 774015

\bibitem[{{Barnab{\`e}} {et~al.}(2013){Barnab{\`e}}, {Spiniello}, {Koopmans},
  {Trager}, {Czoske}, \& {Treu}}]{Bar++13}
{Barnab{\`e}}, M., {Spiniello}, C., {Koopmans}, L.~V.~E., {et~al.} 2013,
  \mnras, 436, 253

\bibitem[{{Bertin} \& {Arnouts}(1996)}]{B+A96}
{Bertin}, E. \& {Arnouts}, S. 1996, \aaps, 117, 393

\bibitem[{{Bolton} {et~al.}(2012){Bolton}, {Brownstein}, {Kochanek}, {Shu},
  {Schlegel}, {Eisenstein}, {Wake}, {Connolly}, {Maraston}, {Arneson}, \&
  {Weaver}}]{Bol++12}
{Bolton}, A.~S., {Brownstein}, J.~R., {Kochanek}, C.~S., {et~al.} 2012, \apj,
  757, 82

\bibitem[{{Bosch} {et~al.}(2018){Bosch}, {Armstrong}, {Bickerton}, {Furusawa},
  {Ikeda}, {Koike}, {Lupton}, {Mineo}, {Price}, {Takata}, {Tanaka}, {Yasuda},
  {AlSayyad}, {Becker}, {Coulton}, {Coupon}, {Garmilla}, {Huang}, {Krughoff},
  {Lang}, {Leauthaud}, {Lim}, {Lust}, {MacArthur}, {Mandelbaum}, {Miyatake},
  {Miyazaki}, {Murata}, {More}, {Okura}, {Owen}, {Swinbank}, {Strauss},
  {Yamada}, \& {Yamanoi}}]{Bos++18}
{Bosch}, J., {Armstrong}, R., {Bickerton}, S., {et~al.} 2018, \pasj, 70, S5

\bibitem[{{Chan} {et~al.}(2020){Chan}, {Suyu}, {Sonnenfeld}, {Jaelani}, {More},
  {Yonehara}, {Kubota}, {Coupon}, {Lee}, {Oguri}, {Rusu}, \& {Wong}}]{Cha++20}
{Chan}, J. H.~H., {Suyu}, S.~H., {Sonnenfeld}, A., {et~al.} 2020, \aap, 636,
  A87

\bibitem[{{Dark Energy Survey Collaboration} {et~al.}(2016){Dark Energy Survey
  Collaboration}, {Abbott}, {Abdalla}, {Aleksi{\'c}}, {Allam}, {Amara},
  {Bacon}, {Balbinot}, {Banerji}, {Bechtol}, {Benoit-L{\'e}vy}, {Bernstein},
  {Bertin}, {Blazek}, {Bonnett}, {Bridle}, {Brooks}, {Brunner}, {Buckley-Geer},
  {Burke}, {Caminha}, {Capozzi}, {Carlsen}, {Carnero-Rosell}, {Carollo},
  {Carrasco-Kind}, {Carretero}, {Castander}, {Clerkin}, {Collett}, {Conselice},
  {Crocce}, {Cunha}, {D'Andrea}, {da Costa}, {Davis}, {Desai}, {Diehl},
  {Dietrich}, {Dodelson}, {Doel}, {Drlica-Wagner}, {Estrada}, {Etherington},
  {Evrard}, {Fabbri}, {Finley}, {Flaugher}, {Foley}, {Fosalba}, {Frieman},
  {Garc{\'\i}a-Bellido}, {Gaztanaga}, {Gerdes}, {Giannantonio}, {Goldstein},
  {Gruen}, {Gruendl}, {Guarnieri}, {Gutierrez}, {Hartley}, {Honscheid}, {Jain},
  {James}, {Jeltema}, {Jouvel}, {Kessler}, {King}, {Kirk}, {Kron}, {Kuehn},
  {Kuropatkin}, {Lahav}, {Li}, {Lima}, {Lin}, {Maia}, {Makler}, {Manera},
  {Maraston}, {Marshall}, {Martini}, {McMahon}, {Melchior}, {Merson}, {Miller},
  {Miquel}, {Mohr}, {Morice-Atkinson}, {Naidoo}, {Neilsen}, {Nichol}, {Nord},
  {Ogando}, {Ostrovski}, {Palmese}, {Papadopoulos}, {Peiris}, {Peoples},
  {Percival}, {Plazas}, {Reed}, {Refregier}, {Romer}, {Roodman}, {Ross},
  {Rozo}, {Rykoff}, {Sadeh}, {Sako}, {S{\'a}nchez}, {Sanchez}, {Santiago},
  {Scarpine}, {Schubnell}, {Sevilla-Noarbe}, {Sheldon}, {Smith}, {Smith},
  {Soares-Santos}, {Sobreira}, {Soumagnac}, {Suchyta}, {Sullivan}, {Swanson},
  {Tarle}, {Thaler}, {Thomas}, {Thomas}, {Tucker}, {Vieira}, {Vikram},
  {Walker}, {Wechsler}, {Weller}, {Wester}, {Whiteway}, {Wilcox}, {Yanny},
  {Zhang}, \& {Zuntz}}]{DES16}
{Dark Energy Survey Collaboration}, {Abbott}, T., {Abdalla}, F.~B., {et~al.}
  2016, \mnras, 460, 1270

\bibitem[{{Davies} {et~al.}(2019){Davies}, {Serjeant}, \& {Bromley}}]{Dav++19}
{Davies}, A., {Serjeant}, S., \& {Bromley}, J.~M. 2019, \mnras, 487, 5263

\bibitem[{{Dawson} {et~al.}(2013){Dawson}, {Schlegel}, {Ahn}, {Anderson},
  {Aubourg}, {Bailey}, {Barkhouser}, {Bautista}, {Beifiori}, {Berlind},
  {Bhardwaj}, {Bizyaev}, {Blake}, {Blanton}, {Blomqvist}, {Bolton}, {Borde},
  {Bovy}, {Brandt}, {Brewington}, {Brinkmann}, {Brown}, {Brownstein}, {Bundy},
  {Busca}, {Carithers}, {Carnero}, {Carr}, {Chen}, {Comparat}, {Connolly},
  {Cope}, {Croft}, {Cuesta}, {da Costa}, {Davenport}, {Delubac}, {de Putter},
  {Dhital}, {Ealet}, {Ebelke}, {Eisenstein}, {Escoffier}, {Fan}, {Filiz Ak},
  {Finley}, {Font-Ribera}, {G{\'e}nova-Santos}, {Gunn}, {Guo}, {Haggard},
  {Hall}, {Hamilton}, {Harris}, {Harris}, {Ho}, {Hogg}, {Holder}, {Honscheid},
  {Huehnerhoff}, {Jordan}, {Jordan}, {Kauffmann}, {Kazin}, {Kirkby}, {Klaene},
  {Kneib}, {Le Goff}, {Lee}, {Long}, {Loomis}, {Lundgren}, {Lupton}, {Maia},
  {Makler}, {Malanushenko}, {Malanushenko}, {Mandelbaum}, {Manera}, {Maraston},
  {Margala}, {Masters}, {McBride}, {McDonald}, {McGreer}, {McMahon}, {Mena},
  {Miralda-Escud{\'e}}, {Montero-Dorta}, {Montesano}, {Muna}, {Myers},
  {Naugle}, {Nichol}, {Noterdaeme}, {Nuza}, {Olmstead}, {Oravetz}, {Oravetz},
  {Owen}, {Padmanabhan}, {Palanque-Delabrouille}, {Pan}, {Parejko},
  {P{\^a}ris}, {Percival}, {P{\'e}rez-Fournon}, {P{\'e}rez-R{\`a}fols},
  {Petitjean}, {Pfaffenberger}, {Pforr}, {Pieri}, {Prada}, {Price-Whelan},
  {Raddick}, {Rebolo}, {Rich}, {Richards}, {Rockosi}, {Roe}, {Ross}, {Ross},
  {Rossi}, {Rubi{\~n}o-Martin}, {Samushia}, {S{\'a}nchez}, {Sayres}, {Schmidt},
  {Schneider}, {Sc{\'o}ccola}, {Seo}, {Shelden}, {Sheldon}, {Shen}, {Shu},
  {Slosar}, {Smee}, {Snedden}, {Stauffer}, {Steele}, {Strauss}, {Streblyanska},
  {Suzuki}, {Swanson}, {Tal}, {Tanaka}, {Thomas}, {Tinker}, {Tojeiro},
  {Tremonti}, {Vargas Maga{\~n}a}, {Verde}, {Viel}, {Wake}, {Watson}, {Weaver},
  {Weinberg}, {Weiner}, {West}, {White}, {Wood-Vasey}, {Yeche}, {Zehavi},
  {Zhao}, \& {Zheng}}]{Daw++13}
{Dawson}, K.~S., {Schlegel}, D.~J., {Ahn}, C.~P., {et~al.} 2013, \aj, 145, 10

\bibitem[{{de Jong} {et~al.}(2015){de Jong}, {Verdoes Kleijn}, {Boxhoorn},
  {Buddelmeijer}, {Capaccioli}, {Getman}, {Grado}, {Helmich}, {Huang},
  {Irisarri}, {Kuijken}, {La Barbera}, {McFarland}, {Napolitano}, {Radovich},
  {Sikkema}, {Valentijn}, {Begeman}, {Brescia}, {Cavuoti}, {Choi}, {Cordes},
  {Covone}, {Dall'Ora}, {Hildebrandt}, {Longo}, {Nakajima}, {Paolillo},
  {Puddu}, {Rifatto}, {Tortora}, {van Uitert}, {Buddendiek},
  {Harnois-D{\'e}raps}, {Erben}, {Eriksen}, {Heymans}, {Hoekstra}, {Joachimi},
  {Kitching}, {Klaes}, {Koopmans}, {K{\"o}hlinger}, {Roy}, {Sif{\'o}n},
  {Schneider}, {Sutherland}, {Viola}, \& {Vriend}}]{deJ++15}
{de Jong}, J. T.~A., {Verdoes Kleijn}, G.~A., {Boxhoorn}, D.~R., {et~al.} 2015,
  \aap, 582, A62

\bibitem[{{Erben} {et~al.}(2013){Erben}, {Hildebrandt}, {Miller}, {van
  Waerbeke}, {Heymans}, {Hoekstra}, {Kitching}, {Mellier}, {Benjamin}, {Blake},
  {Bonnett}, {Cordes}, {Coupon}, {Fu}, {Gavazzi}, {Gillis}, {Grocutt}, {Gwyn},
  {Holhjem}, {Hudson}, {Kilbinger}, {Kuijken}, {Milkeraitis}, {Rowe},
  {Schrabback}, {Semboloni}, {Simon}, {Smit}, {Toader}, {Vafaei}, {van Uitert},
  \& {Velander}}]{Erb++13}
{Erben}, T., {Hildebrandt}, H., {Miller}, L., {et~al.} 2013, \mnras, 433, 2545

\bibitem[{Geach {et~al.}(2015)Geach, More, Verma, Marshall, Jackson, Belles,
  Beswick, Baeten, Chavez, Cornen, Cox, Erben, Erickson, Garrington, Harrison,
  Harrington, Hughes, Ivison, Jordan, Lin, Leauthaud, Lintott, Lynn, Kapadia,
  Kneib, Macmillan, Makler, Miller, Montaña, Mujica, Muxlow, Narayanan,
  Briain, O'Brien, Oguri, Paget, Parrish, Ross, Rozo, Rusu, Rykoff,
  Sanchez-Argüelles, Simpson, Snyder, Schloerb, Tecza, Wang, Van~Waerbeke,
  Wilcox, Viero, Wilson, Yun, \& Zeballos}]{Gea++15}
Geach, J.~E., More, A., Verma, A., {et~al.} 2015, \mnras, 452, 502

\bibitem[{{Grillo} {et~al.}(2018){Grillo}, {Rosati}, {Suyu}, {Balestra},
  {Caminha}, {Halkola}, {Kelly}, {Lombardi}, {Mercurio}, {Rodney}, \&
  {Treu}}]{Gri++18}
{Grillo}, C., {Rosati}, P., {Suyu}, S.~H., {et~al.} 2018, \apj, 860, 94

\bibitem[{{Hildebrandt} {et~al.}(2012){Hildebrandt}, {Erben}, {Kuijken}, {van
  Waerbeke}, {Heymans}, {Coupon}, {Benjamin}, {Bonnett}, {Fu}, {Hoekstra},
  {Kitching}, {Mellier}, {Miller}, {Veland er}, {Hudson}, {Rowe}, {Schrabback},
  {Semboloni}, \& {Ben{\'\i}tez}}]{Hil++12}
{Hildebrandt}, H., {Erben}, T., {Kuijken}, K., {et~al.} 2012, \mnras, 421, 2355

\bibitem[{{Hsueh} {et~al.}(2020){Hsueh}, {Enzi}, {Vegetti}, {Auger},
  {Fassnacht}, {Despali}, {Koopmans}, \& {McKean}}]{Hsu++20}
{Hsueh}, J.~W., {Enzi}, W., {Vegetti}, S., {et~al.} 2020, \mnras, 492, 3047

\bibitem[{{Huang} {et~al.}(2020){Huang}, {Storfer}, {Ravi}, {Pilon}, {Domingo},
  {Schlegel}, {Bailey}, {Dey}, {Gupta}, {Herrera}, {Juneau}, {Landriau},
  {Lang}, {Meisner}, {Moustakas}, {Myers}, {Schlafly}, {Valdes}, {Weaver},
  {Yang}, \& {Y{\`e}che}}]{Hua++20}
{Huang}, X., {Storfer}, C., {Ravi}, V., {et~al.} 2020, \apj, 894, 78

\bibitem[{{Ivezi{\'c}} {et~al.}(2008){Ivezi{\'c}}, {Kahn}, {Tyson}, {Abel},
  {Acosta}, {Allsman}, {Alonso}, {AlSayyad}, {Anderson}, {Andrew}, \&
  et~al.}]{Ive++08}
{Ivezi{\'c}}, {\v Z}., {Kahn}, S.~M., {Tyson}, J.~A., {et~al.} 2008, arXiv
  e-prints [\eprint[arXiv]{0805.2366}]

\bibitem[{{Jacobs} {et~al.}(2019{\natexlab{a}}){Jacobs}, {Collett},
  {Glazebrook}, {Buckley-Geer}, {Diehl}, {Lin}, {McCarthy}, {Qin}, {Odden},
  {Caso Escudero}, {Dial}, {Yung}, {Gaitsch}, {Pellico}, {Lindgren}, {Abbott},
  {Annis}, {Avila}, {Brooks}, {Burke}, {Carnero Rosell}, {Carrasco Kind},
  {Carretero}, {da Costa}, {De Vicente}, {Fosalba}, {Frieman},
  {Garc{\'\i}a-Bellido}, {Gaztanaga}, {Goldstein}, {Gruen}, {Gruendl},
  {Gschwend}, {Hollowood}, {Honscheid}, {Hoyle}, {James}, {Krause},
  {Kuropatkin}, {Lahav}, {Lima}, {Maia}, {Marshall}, {Miquel}, {Plazas},
  {Roodman}, {Sanchez}, {Scarpine}, {Serrano}, {Sevilla-Noarbe}, {Smith},
  {Sobreira}, {Suchyta}, {Swanson}, {Tarle}, {Vikram}, {Walker}, {Zhang}, \&
  {(DES Collaboration}}]{Jac++19}
{Jacobs}, C., {Collett}, T., {Glazebrook}, K., {et~al.} 2019{\natexlab{a}},
  \apjs, 243, 17

\bibitem[{{Jacobs} {et~al.}(2019{\natexlab{b}}){Jacobs}, {Collett},
  {Glazebrook}, {Buckley-Geer}, {Diehl}, {Lin}, {McCarthy}, {Qin}, {Odden},
  {Caso Escudero}, {Dial}, {Yung}, {Gaitsch}, {Pellico}, {Lindgren}, {Abbott},
  {Annis}, {Avila}, {Brooks}, {Burke}, {Carnero Rosell}, {Carrasco Kind},
  {Carretero}, {da Costa}, {De Vicente}, {Fosalba}, {Frieman},
  {Garc{\'\i}a-Bellido}, {Gaztanaga}, {Goldstein}, {Gruen}, {Gruendl},
  {Gschwend}, {Hollowood}, {Honscheid}, {Hoyle}, {James}, {Krause},
  {Kuropatkin}, {Lahav}, {Lima}, {Maia}, {Marshall}, {Miquel}, {Plazas},
  {Roodman}, {Sanchez}, {Scarpine}, {Serrano}, {Sevilla-Noarbe}, {Smith},
  {Sobreira}, {Suchyta}, {Swanson}, {Tarle}, {Vikram}, {Walker}, {Zhang}, \&
  {DES Collaboration}}]{Jac++19b}
{Jacobs}, C., {Collett}, T., {Glazebrook}, K., {et~al.} 2019{\natexlab{b}},
  \apjs, 243, 17

\bibitem[{{Jaelani} {et~al.}(2020{\natexlab{a}}){Jaelani}, {More}, {Oguri},
  {Sonnenfeld}, {Suyu}, {Rusu}, {Wong}, {Chan}, {Kayo}, {Lee}, {Chao},
  {Coupon}, {Inoue}, \& {Futamase}}]{Jae++20b}
{Jaelani}, A.~T., {More}, A., {Oguri}, M., {et~al.} 2020{\natexlab{a}}, \mnras,
  495, 1291

\bibitem[{{Jaelani} {et~al.}(2020{\natexlab{b}}){Jaelani}, {More},
  {Sonnenfeld}, {Oguri}, {Rusu}, {Wong}, {Chan}, {Suyu}, {Kayo}, {Lee}, \&
  {Inoue}}]{Jae++20a}
{Jaelani}, A.~T., {More}, A., {Sonnenfeld}, A., {et~al.} 2020{\natexlab{b}},
  \mnras, 494, 3156

\bibitem[{{Juri{\'c}} {et~al.}(2015){Juri{\'c}}, {Kantor}, {Lim}, {Lupton},
  {Dubois-Felsmann}, {Jenness}, {Axelrod}, {Aleksi{\'c}}, {Allsman},
  {AlSayyad}, {Alt}, {Armstrong}, {Basney}, {Becker}, {Becla}, {Bickerton},
  {Biswas}, {Bosch}, {Boutigny}, {Carrasco Kind}, {Ciardi}, {Connolly},
  {Daniel}, {Daues}, {Economou}, {Chiang}, {Fausti}, {Fisher-Levine},
  {Freemon}, {Gee}, {Gris}, {Hernandez}, {Hoblitt}, {Ivezi{\'c}}, {Jammes},
  {Jevremovi{\'c}}, {Jones}, {Bryce Kalmbach}, {Kasliwal}, {Krughoff}, {Lang},
  {Lurie}, {Lust}, {Mullally}, {MacArthur}, {Melchior}, {Moeyens}, {Nidever},
  {Owen}, {Parejko}, {Peterson}, {Petravick}, {Pietrowicz}, {Price}, {Reiss},
  {Shaw}, {Sick}, {Slater}, {Strauss}, {Sullivan}, {Swinbank}, {Van Dyk},
  {Vuj{\v c}i{\'c}}, {Withers}, {Yoachim}, \& {LSST Project}}]{Jur++15}
{Juri{\'c}}, M., {Kantor}, J., {Lim}, K., {et~al.} 2015, ArXiv e-prints
  [\eprint[arXiv]{1512.07914}]

\bibitem[{{Koopmans} \& {Treu}(2003)}]{K+T03}
{Koopmans}, L.~V.~E. \& {Treu}, T. 2003, \apj, 583, 606

\bibitem[{{Kormann} {et~al.}(1994){Kormann}, {Schneider}, \&
  {Bartelmann}}]{KSB94}
{Kormann}, R., {Schneider}, P., \& {Bartelmann}, M. 1994, \aap, 284, 285

\bibitem[{{Mao} \& {Schneider}(1998)}]{M+S98}
{Mao}, S. \& {Schneider}, P. 1998, \mnras, 295, 587

\bibitem[{{Marshall} {et~al.}(2016){Marshall}, {Verma}, {More}, {Davis},
  {More}, {Kapadia}, {Parrish}, {Snyder}, {Wilcox}, {Baeten}, {Macmillan},
  {Cornen}, {Baumer}, {Simpson}, {Lintott}, {Miller}, {Paget}, {Simpson},
  {Smith}, {K{\"u}ng}, {Saha}, \& {Collett}}]{Mar++16}
{Marshall}, P.~J., {Verma}, A., {More}, A., {et~al.} 2016, \mnras, 455, 1171

\bibitem[{{Mediavilla} {et~al.}(2009){Mediavilla}, {Mu{\~n}oz}, {Falco},
  {Motta}, {Guerras}, {Canovas}, {Jean}, {Oscoz}, \& {Mosquera}}]{Med++09}
{Mediavilla}, E., {Mu{\~n}oz}, J.~A., {Falco}, E., {et~al.} 2009, \apj, 706,
  1451

\bibitem[{{Metcalf} {et~al.}(2019){Metcalf}, {Meneghetti}, {Avestruz},
  {Bellagamba}, {Bom}, {Bertin}, {Cabanac}, {Courbin}, {Davies},
  {Decenci{\`e}re}, {Flamary}, {Gavazzi}, {Geiger}, {Hartley},
  {Huertas-Company}, {Jackson}, {Jacobs}, {Jullo}, {Kneib}, {Koopmans},
  {Lanusse}, {Li}, {Ma}, {Makler}, {Li}, {Lightman}, {Petrillo}, {Serjeant},
  {Sch{\"a}fer}, {Sonnenfeld}, {Tagore}, {Tortora}, {Tuccillo},
  {Valent{\'\i}n}, {Velasco-Forero}, {Verdoes Kleijn}, \&
  {Vernardos}}]{Met++19}
{Metcalf}, R.~B., {Meneghetti}, M., {Avestruz}, C., {et~al.} 2019, \aap, 625,
  A119

\bibitem[{{Millon} {et~al.}(2020){Millon}, {Galan}, {Courbin}, {Treu}, {Suyu},
  {Ding}, {Birrer}, {Chen}, {Shajib}, {Sluse}, {Wong}, {Agnello}, {Auger},
  {Buckley-Geer}, {Chan}, {Collett}, {Fassnacht}, {Hilbert}, {Koopmans},
  {Motta}, {Mukherjee}, {Rusu}, {Sonnenfeld}, {Spiniello}, \& {Van de
  Vyvere}}]{Mil++20}
{Millon}, M., {Galan}, A., {Courbin}, F., {et~al.} 2020, \aap, 639, A101

\bibitem[{{Miyazaki} {et~al.}(2018){Miyazaki}, {Komiyama}, {Kawanomoto}, {Doi},
  {Furusawa}, {Hamana}, {Hayashi}, {Ikeda}, {Kamata}, {Karoji}, {Koike},
  {Kurakami}, {Miyama}, {Morokuma}, {Nakata}, {Namikawa}, {Nakaya}, {Nariai},
  {Obuchi}, {Oishi}, {Okada}, {Okura}, {Tait}, {Takata}, {Tanaka}, {Tanaka},
  {Terai}, {Tomono}, {Uraguchi}, {Usuda}, {Utsumi}, {Yamada}, {Yamanoi},
  {Aihara}, {Fujimori}, {Mineo}, {Miyatake}, {Oguri}, {Uchida}, {Tanaka},
  {Yasuda}, {Takada}, {Murayama}, {Nishizawa}, {Sugiyama}, {Chiba}, {Futamase},
  {Wang}, {Chen}, {Ho}, {Liaw}, {Chiu}, {Ho}, {Lai}, {Lee}, {Jeng}, {Iwamura},
  {Armstrong}, {Bickerton}, {Bosch}, {Gunn}, {Lupton}, {Loomis}, {Price},
  {Smith}, {Strauss}, {Turner}, {Suzuki}, {Miyazaki}, {Muramatsu}, {Yamamoto},
  {Endo}, {Ezaki}, {Ito}, {Kawaguchi}, {Sofuku}, {Taniike}, {Akutsu}, {Dojo},
  {Kasumi}, {Matsuda}, {Imoto}, {Miwa}, {Suzuki}, {Takeshi}, \&
  {Yokota}}]{Miy++18}
{Miyazaki}, S., {Komiyama}, Y., {Kawanomoto}, S., {et~al.} 2018, \pasj, 70, S1

\bibitem[{{More} {et~al.}(2017){More}, {Lee}, {Oguri}, {Ono}, {Suyu}, {Chan},
  {Silverman}, {More}, {Schulze}, {Komiyama}, {Matsuoka}, {Miyazaki}, {Nagao},
  {Ouchi}, {Tait}, {Tanaka}, {Tanaka}, {Usuda}, \& {Yasuda}}]{Mor++17}
{More}, A., {Lee}, C.-H., {Oguri}, M., {et~al.} 2017, \mnras, 465, 2411

\bibitem[{{More} {et~al.}(2009){More}, {McKean}, {More}, {Porcas}, {Koopmans},
  \& {Garrett}}]{Mor++09}
{More}, A., {McKean}, J.~P., {More}, S., {et~al.} 2009, \mnras, 394, 174

\bibitem[{{More} {et~al.}(2016){More}, {Verma}, {Marshall}, {More}, {Baeten},
  {Wilcox}, {Macmillan}, {Cornen}, {Kapadia}, {Parrish}, {Snyder}, {Davis},
  {Gavazzi}, {Lintott}, {Simpson}, {Miller}, {Smith}, {Paget}, {Saha},
  {K{\"u}ng}, \& {Collett}}]{Mor++16}
{More}, A., {Verma}, A., {Marshall}, P.~J., {et~al.} 2016, \mnras, 455, 1191

\bibitem[{{Newman} {et~al.}(2015){Newman}, {Ellis}, \& {Treu}}]{New++15}
{Newman}, A.~B., {Ellis}, R.~S., \& {Treu}, T. 2015, \apj, 814, 26

\bibitem[{{Oguri} {et~al.}(2014){Oguri}, {Rusu}, \& {Falco}}]{ORF14}
{Oguri}, M., {Rusu}, C.~E., \& {Falco}, E.~E. 2014, \mnras, 439, 2494

\bibitem[{{Oldham} \& {Auger}(2018)}]{O+A18}
{Oldham}, L.~J. \& {Auger}, M.~W. 2018, \mnras, 476, 133

\bibitem[{{Petrillo} {et~al.}(2019{\natexlab{a}}){Petrillo}, {Tortora},
  {Vernardos}, {Koopmans}, {Verdoes Kleijn}, {Bilicki}, {Napolitano},
  {Chatterjee}, {Covone}, {Dvornik}, {Erben}, {Getman}, {Giblin}, {Heymans},
  {de Jong}, {Kuijken}, {Schneider}, {Shan}, {Spiniello}, \&
  {Wright}}]{Pet++19}
{Petrillo}, C.~E., {Tortora}, C., {Vernardos}, G., {et~al.} 2019{\natexlab{a}},
  \mnras, 484, 3879

\bibitem[{{Petrillo} {et~al.}(2019{\natexlab{b}}){Petrillo}, {Tortora},
  {Vernardos}, {Koopmans}, {Verdoes Kleijn}, {Bilicki}, {Napolitano},
  {Chatterjee}, {Covone}, {Dvornik}, {Erben}, {Getman}, {Giblin}, {Heymans},
  {de Jong}, {Kuijken}, {Schneider}, {Shan}, {Spiniello}, \&
  {Wright}}]{Pet++19b}
{Petrillo}, C.~E., {Tortora}, C., {Vernardos}, G., {et~al.} 2019{\natexlab{b}},
  \mnras, 484, 3879

\bibitem[{{Ruff} {et~al.}(2011){Ruff}, {Gavazzi}, {Marshall}, {Treu}, {Auger},
  \& {Brault}}]{Ruf++11}
{Ruff}, A.~J., {Gavazzi}, R., {Marshall}, P.~J., {et~al.} 2011, \apj, 727, 96

\bibitem[{{Schechter} {et~al.}(2014){Schechter}, {Pooley}, {Blackburne}, \&
  {Wambsganss}}]{Sch++14}
{Schechter}, P.~L., {Pooley}, D., {Blackburne}, J.~A., \& {Wambsganss}, J.
  2014, \apj, 793, 96

\bibitem[{{Smith} {et~al.}(2020){Smith}, {Collier}, {Ozaki}, \&
  {Lucey}}]{Smi++20}
{Smith}, R.~J., {Collier}, W.~P., {Ozaki}, S., \& {Lucey}, J.~R. 2020, \mnras,
  493, L33

\bibitem[{{Smith} {et~al.}(2018){Smith}, {Lucey}, \& {Collier}}]{SLC18}
{Smith}, R.~J., {Lucey}, J.~R., \& {Collier}, W.~P. 2018, \mnras, 481, 2115

\bibitem[{{Smith} {et~al.}(2015){Smith}, {Lucey}, \& {Conroy}}]{SLC15}
{Smith}, R.~J., {Lucey}, J.~R., \& {Conroy}, C. 2015, \mnras, 449, 3441

\bibitem[{{Sonnenfeld} {et~al.}(2018){Sonnenfeld}, {Chan}, {Shu}, {More},
  {Oguri}, {Suyu}, {Wong}, {Lee}, {Coupon}, {Yonehara}, {Bolton}, {Jaelani},
  {Tanaka}, {Miyazaki}, \& {Komiyama}}]{Son++18}
{Sonnenfeld}, A., {Chan}, J.~H.~H., {Shu}, Y., {et~al.} 2018, \pasj, 70, S29

\bibitem[{{Sonnenfeld} {et~al.}(2013{\natexlab{a}}){Sonnenfeld}, {Gavazzi},
  {Suyu}, {Treu}, \& {Marshall}}]{Son++13a}
{Sonnenfeld}, A., {Gavazzi}, R., {Suyu}, S.~H., {Treu}, T., \& {Marshall},
  P.~J. 2013{\natexlab{a}}, \apj, 777, 97

\bibitem[{{Sonnenfeld} {et~al.}(2019){Sonnenfeld}, {Jaelani}, {Chan}, {More},
  {Suyu}, {Wong}, {Oguri}, \& {Lee}}]{Son++19b}
{Sonnenfeld}, A., {Jaelani}, A.~T., {Chan}, J., {et~al.} 2019, \aap, 630, A71

\bibitem[{{Sonnenfeld} {et~al.}(2012){Sonnenfeld}, {Treu}, {Gavazzi},
  {Marshall}, {Auger}, {Suyu}, {Koopmans}, \& {Bolton}}]{Son++12}
{Sonnenfeld}, A., {Treu}, T., {Gavazzi}, R., {et~al.} 2012, \apj, 752, 163

\bibitem[{{Sonnenfeld} {et~al.}(2013{\natexlab{b}}){Sonnenfeld}, {Treu},
  {Gavazzi}, {Suyu}, {Marshall}, {Auger}, \& {Nipoti}}]{Son++13b}
{Sonnenfeld}, A., {Treu}, T., {Gavazzi}, R., {et~al.} 2013{\natexlab{b}}, \apj,
  777, 98

\bibitem[{{Sonnenfeld} {et~al.}(2015){Sonnenfeld}, {Treu}, {Marshall}, {Suyu},
  {Gavazzi}, {Auger}, \& {Nipoti}}]{Son++15}
{Sonnenfeld}, A., {Treu}, T., {Marshall}, P.~J., {et~al.} 2015, \apj, 800, 94

\bibitem[{{Spiniello} {et~al.}(2012){Spiniello}, {Trager}, {Koopmans}, \&
  {Chen}}]{Spi++12}
{Spiniello}, C., {Trager}, S.~C., {Koopmans}, L.~V.~E., \& {Chen}, Y.~P. 2012,
  \apjl, 753, L32

\bibitem[{{Suyu} {et~al.}(2017){Suyu}, {Bonvin}, {Courbin}, {Fassnacht},
  {Rusu}, {Sluse}, {Treu}, {Wong}, {Auger}, {Ding}, {Hilbert}, {Marshall},
  {Rumbaugh}, {Sonnenfeld}, {Tewes}, {Tihhonova}, {Agnello}, {Blandford},
  {Chen}, {Collett}, {Koopmans}, {Liao}, {Meylan}, \& {Spiniello}}]{Suy++17}
{Suyu}, S.~H., {Bonvin}, V., {Courbin}, F., {et~al.} 2017, \mnras, 468, 2590

\bibitem[{{Suyu} {et~al.}(2012){Suyu}, {Hensel}, {McKean}, {Fassnacht}, {Treu},
  {Halkola}, {Norbury}, {Jackson}, {Schneider}, {Thompson}, {Auger},
  {Koopmans}, \& {Matthews}}]{Suy++12}
{Suyu}, S.~H., {Hensel}, S.~W., {McKean}, J.~P., {et~al.} 2012, \apj, 750, 10

\bibitem[{{Tanaka}(2015)}]{Tan++15}
{Tanaka}, M. 2015, \apj, 801, 20

\bibitem[{{Tanaka} {et~al.}(2018){Tanaka}, {Coupon}, {Hsieh}, {Mineo},
  {Nishizawa}, {Speagle}, {Furusawa}, {Miyazaki}, \& {Murayama}}]{Tan++18}
{Tanaka}, M., {Coupon}, J., {Hsieh}, B.-C., {et~al.} 2018, \pasj, 70, S9

\bibitem[{{Treu} {et~al.}(2010){Treu}, {Auger}, {Koopmans}, {Gavazzi},
  {Marshall}, \& {Bolton}}]{Tre++10}
{Treu}, T., {Auger}, M.~W., {Koopmans}, L.~V.~E., {et~al.} 2010, \apj, 709,
  1195

\bibitem[{{Treu} {et~al.}(2011){Treu}, {Dutton}, {Auger}, {Marshall}, {Bolton},
  {Brewer}, {Koo}, \& {Koopmans}}]{Tre++11}
{Treu}, T., {Dutton}, A.~A., {Auger}, M.~W., {et~al.} 2011, \mnras, 417, 1601

\bibitem[{{Treu} \& {Koopmans}(2002)}]{T+K02}
{Treu}, T. \& {Koopmans}, L.~V.~E. 2002, \apj, 575, 87

\bibitem[{{Vegetti} {et~al.}(2010){Vegetti}, {Koopmans}, {Bolton}, {Treu}, \&
  {Gavazzi}}]{Veg++10}
{Vegetti}, S., {Koopmans}, L.~V.~E., {Bolton}, A., {Treu}, T., \& {Gavazzi}, R.
  2010, \mnras, 408, 1969

\bibitem[{{Wong} {et~al.}(2018){Wong}, {Sonnenfeld}, {Chan}, {Rusu}, {Tanaka},
  {Jaelani}, {Lee}, {More}, {Oguri}, \& {Suyu}}]{Won++18}
{Wong}, K.~C., {Sonnenfeld}, A., {Chan}, J. H.~H., {et~al.} 2018, \apj, 867,
  107

\bibitem[{{Wong} {et~al.}(2020){Wong}, {Suyu}, {Chen}, {Rusu}, {Millon},
  {Sluse}, {Bonvin}, {Fassnacht}, {Taubenberger}, {Auger}, {Birrer}, {Chan},
  {Courbin}, {Hilbert}, {Tihhonova}, {Treu}, {Agnello}, {Ding}, {Jee},
  {Komatsu}, {Shajib}, {Sonnenfeld}, {Bland ford}, {Koopmans}, {Marshall}, \&
  {Meylan}}]{Won++20}
{Wong}, K.~C., {Suyu}, S.~H., {Chen}, G. C.~F., {et~al.} 2020, \mnras
  [\eprint[arXiv]{1907.04869}]

\end{thebibliography}

\clearpage
\onecolumn

\begin{longtable}{lcccccccc}
\caption{\label{tab:gradeAB} Grade A and B lens candidates. Columns 7 and 8 indicate whether the candidate was found by \yatta\ or was noted from the `Talk' section of the Space Warps website. Column 9 lists references to papers that include the same lens candidate (and that were published after the beginning of the Space Warps~-~HSC campaign), as follows: $^1$\citet{Jae++20b}, $^2$\citet{Cha++20}, $^3$\citet{Pet++19}, $^4$\citet{Jac++19}, $^5$\citet{Hua++20}}\\
\hline\hline
Name & R.A. & Dec. & $z_{\mathrm{phot}}$ & Grade & $\pr(\mathrm{LENS}|\data)$ & YL & Talk& References \\
 & (deg) & (deg) & & & & & & \\
\hline
\endfirsthead
\caption{continued.}\\
\hline\hline
Name & R.A. & Dec. & $z_{\mathrm{phot}}$ & Grade & $\pr(\mathrm{LENS}|\data)$ & YL & Talk& References \\
 & (deg) & (deg) & & & & & & \\
\hline
\endhead
\hline
\endfoot
HSCJ015913.34$-$054320.7 & $29.8056$ & $-5.7224$ & $0.93$ & B & $1.00$ & N & N & $\ldots$\\
HSCJ015938.21$-$035859.0 & $29.9092$ & $-3.9831$ & $1.16$ & B & $0.80$ & N & N & $^{1}$\\
HSCJ020018.51$-$030549.6 & $30.0771$ & $-3.0971$ & $0.77$ & B & $1.00$ & Y & Y & $\ldots$\\
HSCJ020050.32$-$030027.1 & $30.2097$ & $-3.0075$ & $0.73$ & B & $1.00$ & N & Y & $\ldots$\\
HSCJ020449.88$-$020206.2 & $31.2078$ & $-2.0350$ & $0.85$ & B & $1.00$ & N & Y & $\ldots$\\
HSCJ020810.69$-$022018.2 & $32.0445$ & $-2.3384$ & $0.63$ & B & $1.00$ & Y & Y & $\ldots$\\
HSCJ020816.10$-$023724.1 & $32.0671$ & $-2.6234$ & $0.48$ & B & $1.00$ & N & N & $^{1}$\\
HSCJ020955.42$-$024442.2 & $32.4809$ & $-2.7451$ & $0.72$ & B & $1.00$ & Y & Y & $^{1}$\\
HSCJ021134.26$-$023752.4 & $32.8928$ & $-2.6312$ & $0.47$ & B & $4.00\times 10^{-2}$ & N & Y & $\ldots$\\
HSCJ021408.00$-$020628.4 & $33.5333$ & $-2.1079$ & $0.70$ & B & $1.00$ & N & Y & $^{1, 4}$\\
HSCJ021645.36$-$020835.2 & $34.1890$ & $-2.1431$ & $0.61$ & B & $1.00$ & N & Y & $\ldots$\\
HSCJ021829.18$-$034040.8 & $34.6216$ & $-3.6780$ & $0.89$ & B & $1.00$ & Y & Y & $\ldots$\\
HSCJ022422.34$-$014946.5 & $36.0931$ & $-1.8296$ & $0.61$ & B & $1.00$ & Y & Y & $\ldots$\\
HSCJ023038.13$-$015248.9 & $37.6589$ & $-1.8803$ & $0.40$ & B & $1.00$ & N & Y & $\ldots$\\
HSCJ023133.59$-$042501.6 & $37.8900$ & $-4.4171$ & $0.92$ & B & $1.80\times 10^{-4}$ & N & Y & $\ldots$\\
HSCJ023150.80$-$020509.2 & $37.9617$ & $-2.0859$ & $0.76$ & B & $1.00$ & N & Y & $^{1}$\\
HSCJ023305.59$-$022836.2 & $38.2733$ & $-2.4767$ & $0.58$ & A & $1.00$ & Y & N & $^{1}$\\
HSCJ023322.66$-$020530.4 & $38.3444$ & $-2.0918$ & $0.74$ & A & $0.91$ & Y & Y & $^{1}$\\
HSCJ023331.95$-$032801.1 & $38.3831$ & $-3.4670$ & $1.12$ & B & $0.98$ & Y & N & $^{1}$\\
HSCJ023629.19$-$060449.4 & $39.1216$ & $-6.0804$ & $0.91$ & B & $1.00$ & N & Y & $\ldots$\\
HSCJ084211.15$+$013835.7 & $130.5465$ & $1.6432$ & $0.28$ & B & $1.00$ & Y & N & $^{3}$\\
HSCJ084520.15$-$005456.3 & $131.3340$ & $-0.9156$ & $0.34$ & A & $1.00$ & N & Y & $^{1, 3}$\\
HSCJ084536.06$-$000456.9 & $131.4002$ & $-0.0825$ & $0.81$ & B & $1.00$ & Y & Y & $\ldots$\\
HSCJ084958.90$-$005306.2 & $132.4954$ & $-0.8851$ & $0.47$ & B & $1.00$ & N & Y & $\ldots$\\
HSCJ085046.61$+$003905.4 & $132.6942$ & $0.6515$ & $0.84$ & A & $1.00$ & Y & Y & $\ldots$\\
HSCJ085246.78$+$021959.8 & $133.1949$ & $2.3333$ & $0.68$ & B & $1.00$ & Y & N & $\ldots$\\
HSCJ085842.33$-$002745.7 & $134.6764$ & $-0.4627$ & $0.61$ & B & $1.00$ & N & Y & $^{1}$\\
HSCJ085947.63$-$003951.7 & $134.9484$ & $-0.6644$ & $0.56$ & B & $1.00$ & N & Y & $\ldots$\\
HSCJ090146.27$+$040732.2 & $135.4428$ & $4.1256$ & $0.67$ & B & $1.00$ & N & N & $\ldots$\\
HSCJ090241.10$+$025318.2 & $135.6713$ & $2.8884$ & $0.92$ & B & $1.00$ & Y & Y & $\ldots$\\
HSCJ090404.33$+$012516.1 & $136.0180$ & $1.4211$ & $0.82$ & B & $1.00$ & N & Y & $^{1}$\\
HSCJ090502.49$+$004424.0 & $136.2604$ & $0.7400$ & $0.74$ & B & $1.00$ & Y & Y & $\ldots$\\
HSCJ090548.79$+$004743.4 & $136.4533$ & $0.7954$ & $0.93$ & B & $1.00$ & N & N & $\ldots$\\
HSCJ090611.05$+$011951.7 & $136.5460$ & $1.3310$ & $0.65$ & B & $1.00$ & N & Y & $^{1}$\\
HSCJ090618.93$+$003053.8 & $136.5789$ & $0.5150$ & $0.88$ & B & $1.00$ & Y & N & $\ldots$\\
HSCJ090707.09$-$004741.5 & $136.7795$ & $-0.7948$ & $0.76$ & B & $0.72$ & N & N & $\ldots$\\
HSCJ090754.35$+$005732.3 & $136.9765$ & $0.9590$ & $0.68$ & A & $1.00$ & N & Y & $^{1}$\\
HSCJ090806.26$+$011955.6 & $137.0261$ & $1.3321$ & $0.65$ & B & $1.00$ & Y & Y & $^{1}$\\
HSCJ090822.32$-$010752.3 & $137.0930$ & $-1.1312$ & $0.77$ & B & $1.00$ & Y & Y & $\ldots$\\
HSCJ090938.60$+$002842.6 & $137.4108$ & $0.4785$ & $0.74$ & B & $1.00$ & Y & Y & $\ldots$\\
HSCJ091031.61$+$003142.9 & $137.6317$ & $0.5286$ & $0.25$ & B & $1.00$ & Y & Y & $\ldots$\\
HSCJ091033.14$+$000712.1 & $137.6381$ & $0.1200$ & $0.48$ & B & $0.77$ & N & N & $\ldots$\\
HSCJ091148.92$+$041852.9 & $137.9538$ & $4.3147$ & $0.81$ & B & $0.98$ & Y & Y & $\ldots$\\
HSCJ091351.92$+$014131.4 & $138.4663$ & $1.6921$ & $0.47$ & B & $1.00$ & Y & Y & $\ldots$\\
HSCJ091413.47$-$002444.1 & $138.5561$ & $-0.4123$ & $0.75$ & B & $1.00$ & Y & N & $\ldots$\\
HSCJ091740.87$+$004438.1 & $139.4203$ & $0.7439$ & $0.31$ & B & $1.00$ & Y & N & $\ldots$\\
HSCJ091843.38$-$022007.3 & $139.6808$ & $-2.3354$ & $0.59$ & B & $1.10\times 10^{-5}$ & N & Y & $\ldots$\\
HSCJ092120.76$+$044430.6 & $140.3365$ & $4.7418$ & $0.59$ & B & $1.00$ & N & Y & $\ldots$\\
HSCJ092136.66$+$021409.5 & $140.4028$ & $2.2360$ & $0.34$ & B & $1.00$ & Y & Y & $^{1}$\\
HSCJ092545.00$+$001702.8 & $141.4375$ & $0.2841$ & $0.76$ & B & $1.00$ & Y & Y & $^{1}$\\
HSCJ114233.36$+$004607.0 & $175.6390$ & $0.7686$ & $0.61$ & B & $1.00$ & N & Y & $\ldots$\\
HSCJ114438.85$-$002547.1 & $176.1619$ & $-0.4297$ & $0.70$ & B & $1.00$ & Y & Y & $^{1}$\\
HSCJ114444.97$+$001344.9 & $176.1874$ & $0.2291$ & $0.92$ & B & $1.00$ & N & Y & $\ldots$\\
HSCJ114445.56$+$005256.5 & $176.1898$ & $0.8824$ & $0.72$ & B & $1.00$ & N & N & $\ldots$\\
HSCJ115011.26$-$002019.8 & $177.5469$ & $-0.3388$ & $0.65$ & B & $1.00$ & Y & Y & $\ldots$\\
HSCJ115057.41$-$015316.8 & $177.7392$ & $-1.8880$ & $0.94$ & B & $5.00\times 10^{-6}$ & N & Y & $\ldots$\\
HSCJ115315.60$+$012746.0 & $178.3150$ & $1.4628$ & $0.47$ & A & $1.00$ & N & Y & $^{1}$\\
HSCJ115529.44$-$004255.9 & $178.8727$ & $-0.7155$ & $0.82$ & B & $1.00$ & N & N & $\ldots$\\
HSCJ115630.95$-$020027.6 & $179.1289$ & $-2.0077$ & $1.15$ & B & $1.00\times 10^{-6}$ & N & Y & $\ldots$\\
HSCJ120053.28$-$013357.8 & $180.2220$ & $-1.5661$ & $0.92$ & B & $1.00$ & Y & Y & $\ldots$\\
HSCJ120109.87$+$003330.0 & $180.2911$ & $0.5583$ & $1.11$ & B & $1.00$ & Y & Y & $\ldots$\\
HSCJ120111.51$+$012635.0 & $180.2980$ & $1.4431$ & $0.59$ & B & $1.00$ & N & Y & $\ldots$\\
HSCJ120129.19$+$001930.0 & $180.3716$ & $0.3250$ & $0.72$ & B & $1.00$ & Y & Y & $\ldots$\\
HSCJ120141.96$+$012737.0 & $180.4248$ & $1.4603$ & $0.75$ & B & $0.16$ & N & Y & $\ldots$\\
HSCJ120256.87$+$003930.5 & $180.7369$ & $0.6585$ & $0.72$ & A & $1.00$ & N & N & $^{1}$\\
HSCJ120630.99$-$011428.7 & $181.6291$ & $-1.2413$ & $0.69$ & B & $1.00$ & N & Y & $\ldots$\\
HSCJ120806.60$-$012233.2 & $182.0275$ & $-1.3759$ & $0.36$ & B & $1.00$ & Y & N & $\ldots$\\
HSCJ120849.26$+$012845.2 & $182.2052$ & $1.4792$ & $0.99$ & B & $1.00$ & Y & Y & $^{1}$\\
HSCJ120855.31$-$010304.7 & $182.2305$ & $-1.0513$ & $0.64$ & B & $0.93$ & N & N & $^{1}$\\
HSCJ135942.84$-$012620.9 & $209.9285$ & $-1.4391$ & $0.33$ & B & $1.00$ & Y & Y & $^{1}$\\
HSCJ140042.84$-$010556.7 & $210.1785$ & $-1.0991$ & $0.70$ & B & $1.00$ & Y & Y & $\ldots$\\
HSCJ140452.30$+$005122.6 & $211.2179$ & $0.8563$ & $0.33$ & B & $1.00$ & Y & N & $^{3}$\\
HSCJ140753.62$-$002816.6 & $211.9734$ & $-0.4713$ & $0.44$ & B & $1.00$ & N & Y & $^{1}$\\
HSCJ141050.10$-$010938.6 & $212.7088$ & $-1.1607$ & $0.63$ & B & $1.00$ & N & Y & $^{1}$\\
HSCJ141105.77$+$002532.5 & $212.7741$ & $0.4257$ & $1.13$ & B & $1.00$ & N & N & $^{1}$\\
HSCJ141136.52$-$010215.6 & $212.9022$ & $-1.0377$ & $0.94$ & A & $1.00$ & Y & Y & $\ldots$\\
HSCJ141137.75$+$010720.4 & $212.9073$ & $1.1223$ & $0.44$ & B & $1.00$ & Y & Y & $^{1, 3}$\\
HSCJ141420.98$-$013646.2 & $213.5874$ & $-1.6128$ & $0.49$ & A & $1.00$ & N & Y & $^{3}$\\
HSCJ141435.59$-$001601.7 & $213.6483$ & $-0.2671$ & $0.71$ & B & $1.00$ & N & N & $\ldots$\\
HSCJ141558.21$+$523955.4 & $213.9926$ & $52.6654$ & $0.45$ & B & $1.00$ & Y & N & $\ldots$\\
HSCJ141646.59$-$011813.3 & $214.1941$ & $-1.3037$ & $0.66$ & B & $1.00$ & N & N & $\ldots$\\
HSCJ141649.82$+$013822.3 & $214.2076$ & $1.6395$ & $0.33$ & B & $1.00$ & Y & Y & $\ldots$\\
HSCJ141805.58$+$004435.4 & $214.5232$ & $0.7432$ & $0.85$ & B & $1.00$ & N & Y & $^{1}$\\
HSCJ141908.45$+$002049.9 & $214.7852$ & $0.3472$ & $0.25$ & B & $1.00$ & N & Y & $^{1}$\\
HSCJ141933.31$+$010223.8 & $214.8888$ & $1.0399$ & $1.12$ & B & $1.00$ & Y & N & $\ldots$\\
HSCJ142017.81$+$005832.0 & $215.0742$ & $0.9755$ & $0.25$ & B & $1.00$ & Y & N & $^{1}$\\
HSCJ142018.01$+$005832.9 & $215.0751$ & $0.9758$ & $0.78$ & B & $0.69$ & N & N & $\ldots$\\
HSCJ142031.47$+$002021.1 & $215.1311$ & $0.3392$ & $0.46$ & B & $1.00$ & N & Y & $\ldots$\\
HSCJ142103.70$+$002219.2 & $215.2654$ & $0.3720$ & $0.64$ & B & $1.00$ & Y & N & $^{1, 5}$\\
HSCJ142149.87$-$002403.0 & $215.4578$ & $-0.4008$ & $0.58$ & B & $1.00$ & N & N & $^{1}$\\
HSCJ142232.32$+$000134.6 & $215.6347$ & $0.0263$ & $0.23$ & B & $1.00$ & Y & Y & $^{3}$\\
HSCJ142234.29$-$000225.4 & $215.6429$ & $-0.0404$ & $1.05$ & B & $1.00$ & Y & N & $\ldots$\\
HSCJ142336.96$-$004034.7 & $215.9040$ & $-0.6763$ & $0.88$ & B & $1.00$ & Y & N & $\ldots$\\
HSCJ142602.93$+$010842.0 & $216.5122$ & $1.1450$ & $0.86$ & B & $1.00$ & N & N & $\ldots$\\
HSCJ142640.64$+$000958.6 & $216.6693$ & $0.1663$ & $0.59$ & B & $1.00$ & N & N & $\ldots$\\
HSCJ142754.88$+$003944.6 & $216.9787$ & $0.6624$ & $0.82$ & B & $1.00$ & N & Y & $\ldots$\\
HSCJ142811.66$-$005021.2 & $217.0486$ & $-0.8392$ & $0.65$ & B & $1.00$ & Y & Y & $\ldots$\\
HSCJ143113.89$-$000613.5 & $217.8079$ & $-0.1037$ & $0.73$ & B & $1.00$ & Y & Y & $^{1}$\\
HSCJ143150.86$+$013019.9 & $217.9619$ & $1.5055$ & $0.72$ & B & $1.00$ & Y & Y & $\ldots$\\
HSCJ143243.95$-$004553.4 & $218.1831$ & $-0.7648$ & $0.45$ & B & $1.00$ & Y & Y & $\ldots$\\
HSCJ143408.93$-$001311.7 & $218.5372$ & $-0.2199$ & $0.88$ & B & $1.00$ & Y & N & $\ldots$\\
HSCJ143518.88$-$012526.2 & $218.8287$ & $-1.4239$ & $0.98$ & B & $0.89$ & N & N & $\ldots$\\
HSCJ143519.11$-$010636.2 & $218.8296$ & $-1.1101$ & $0.76$ & B & $1.00$ & Y & Y & $^{1}$\\
HSCJ143529.13$+$001812.5 & $218.8714$ & $0.3035$ & $0.60$ & B & $1.00$ & Y & N & $\ldots$\\
HSCJ143901.39$+$005117.9 & $219.7558$ & $0.8550$ & $0.75$ & B & $1.00$ & Y & Y & $\ldots$\\
HSCJ143917.26$-$003917.9 & $219.8219$ & $-0.6550$ & $0.71$ & B & $1.00$ & N & Y & $\ldots$\\
HSCJ144101.48$+$013344.8 & $220.2562$ & $1.5624$ & $0.53$ & B & $1.00$ & Y & Y & $\ldots$\\
HSCJ144132.68$-$005358.2 & $220.3862$ & $-0.8995$ & $0.49$ & A & $1.00$ & N & N & $\ldots$\\
HSCJ144230.97$-$002353.2 & $220.6291$ & $-0.3981$ & $0.40$ & B & $1.00$ & Y & Y & $^{2, 3}$\\
HSCJ144320.64$-$012537.1 & $220.8360$ & $-1.4270$ & $1.16$ & A & $1.00$ & Y & Y & $^{2}$\\
HSCJ144603.49$-$004016.6 & $221.5146$ & $-0.6713$ & $0.23$ & B & $1.00$ & Y & N & $^{3}$\\
HSCJ145020.04$+$003535.1 & $222.5835$ & $0.5931$ & $1.15$ & B & $4.50\times 10^{-5}$ & N & Y & $\ldots$\\
HSCJ145217.42$-$014714.3 & $223.0726$ & $-1.7873$ & $0.47$ & B & $1.00$ & Y & N & $\ldots$\\
HSCJ150021.02$-$004936.8 & $225.0876$ & $-0.8269$ & $0.41$ & A & $1.00$ & N & Y & $\ldots$\\
HSCJ155358.98$+$425933.5 & $238.4957$ & $42.9926$ & $0.67$ & B & $1.00$ & N & Y & $\ldots$\\
HSCJ155619.50$+$422854.8 & $239.0812$ & $42.4819$ & $1.01$ & B & $1.00$ & N & N & $\ldots$\\
HSCJ155634.86$+$432413.8 & $239.1452$ & $43.4038$ & $0.69$ & B & $1.00$ & N & Y & $\ldots$\\
HSCJ155806.14$+$431702.3 & $239.5256$ & $43.2840$ & $0.75$ & B & $1.00$ & N & Y & $\ldots$\\
HSCJ155814.67$+$423707.8 & $239.5611$ & $42.6188$ & $0.76$ & B & $1.00$ & Y & Y & $\ldots$\\
HSCJ155830.18$+$443339.7 & $239.6258$ & $44.5610$ & $0.58$ & B & $1.00$ & Y & N & $\ldots$\\
HSCJ155950.37$+$424950.5 & $239.9599$ & $42.8307$ & $0.76$ & B & $1.00$ & N & N & $\ldots$\\
HSCJ161022.20$+$422228.9 & $242.5925$ & $42.3747$ & $0.64$ & B & $0.98$ & Y & N & $\ldots$\\
HSCJ161432.69$+$435125.2 & $243.6362$ & $43.8570$ & $0.92$ & B & $0.87$ & N & Y & $\ldots$\\
HSCJ161818.58$+$434527.4 & $244.5774$ & $43.7576$ & $0.69$ & B & $0.63$ & N & Y & $^{1}$\\
HSCJ162422.32$+$433958.0 & $246.0930$ & $43.6661$ & $0.92$ & B & $1.00$ & Y & N & $\ldots$\\
HSCJ221222.80$-$001811.3 & $333.0950$ & $-0.3031$ & $0.93$ & B & $1.00$ & Y & Y & $^{1}$\\
HSCJ222041.41$+$004912.4 & $335.1725$ & $0.8201$ & $0.81$ & B & $1.00$ & N & Y & $\ldots$\\
HSCJ222347.40$+$004928.8 & $335.9475$ & $0.8247$ & $1.16$ & B & $7.00\times 10^{-6}$ & N & Y & $\ldots$\\
HSCJ222445.39$+$010916.8 & $336.1891$ & $1.1547$ & $0.76$ & B & $9.00\times 10^{-6}$ & N & Y & $\ldots$\\
HSCJ222515.86$+$004822.6 & $336.3161$ & $0.8063$ & $0.28$ & B & $1.00$ & Y & N & $\ldots$\\
HSCJ222638.29$-$003449.4 & $336.6596$ & $-0.5804$ & $0.36$ & A & $1.00$ & Y & Y & $^{1}$\\
HSCJ222723.16$+$003257.7 & $336.8465$ & $0.5494$ & $0.85$ & B & $1.00$ & Y & Y & $\ldots$\\
HSCJ223032.20$-$003922.1 & $337.6342$ & $-0.6561$ & $0.35$ & B & $0.69$ & N & N & $\ldots$\\
HSCJ223200.29$+$004015.2 & $338.0012$ & $0.6709$ & $0.90$ & B & $1.00$ & Y & N & $\ldots$\\
HSCJ223443.93$+$030307.9 & $338.6830$ & $3.0522$ & $0.57$ & B & $1.00$ & Y & Y & $\ldots$\\
HSCJ223626.31$-$005400.7 & $339.1096$ & $-0.9002$ & $0.99$ & B & $1.00$ & Y & Y & $\ldots$\\
HSCJ223735.36$+$004014.8 & $339.3974$ & $0.6708$ & $0.47$ & B & $1.00$ & Y & N & $\ldots$\\
HSCJ223934.68$+$023507.1 & $339.8945$ & $2.5853$ & $0.91$ & A & $1.00$ & N & Y & $^{1}$\\
HSCJ224154.63$+$000331.6 & $340.4776$ & $0.0588$ & $0.58$ & B & $1.00$ & Y & Y & $\ldots$\\
HSCJ224211.63$+$023023.8 & $340.5485$ & $2.5066$ & $0.75$ & B & $2.90\times 10^{-4}$ & N & Y & $\ldots$\\
HSCJ224306.92$-$010750.9 & $340.7789$ & $-1.1308$ & $0.48$ & B & $1.00$ & Y & Y & $\ldots$\\
HSCJ224429.44$-$001759.3 & $341.1226$ & $-0.2998$ & $0.85$ & B & $1.00$ & Y & Y & $\ldots$\\
HSCJ224758.49$+$020648.9 & $341.9937$ & $2.1136$ & $0.72$ & B & $0.37$ & N & Y & $\ldots$\\
\end{longtable}

\end{document}